\documentclass[12pt]{article}
\usepackage{amsmath,amssymb,exscale,color}
\usepackage{epsfig,psfrag}
\usepackage{color,amsmath,amscd,shadow,fancybox,axodraw,booktabs}
\newcommand{\Dslash}{{\not \!\!D}}
\newcommand{\Aslash}{{\not \!\!A}}

\begin{document}
\topmargin -1.0cm
\oddsidemargin -0.8cm
\evensidemargin -0.8cm

\thispagestyle{empty}

\vspace{20pt}

\begin{center}
\vspace{20pt}

\Large \textbf{Solving the $A^b_{FB}$ anomaly in natural composite models}

\end{center}

\vspace{15pt}
\begin{center}
{\large  Leandro Da Rold} 

\vspace{20pt}

\textit{Centro At\'{o}mico Bariloche, 8400 San Carlos de Bariloche, Argentina}

\end{center}

\vspace{20pt}
\begin{center}
\textbf{Abstract}
\end{center}
\vspace{5pt} {\small \noindent
The Standard Model with a light Higgs provides a very accurate description of the electroweak precision observables. The largest deviation between the Standard Model predictions and the experimental measurements, the forward-backward asymmetry of the bottom quark $A^b_{FB}$, can be interpreted as an indication of new physics at the TeV scale. The strong agreement between theory and experiment in the branching fraction of the $Z$ into $b$-quarks puts strong constraints for new physics aiming to solve the $A^b_{FB}$ puzzle. We study a class of natural composite Higgs models that can solve the $A^b_{FB}$ anomaly reproducing the observed $R_b$ as well as the top and bottom masses. We find that the subgroup of the custodial symmetry able to protect $Zb_L\bar b_L$ from large corrections generated by the top sector play an important role if we want to maintain naturalness in the composite sector. We make a thorough study of the composite operators mixing with the $b$-quark, determine their embedding under the global composite symmetry and the parameter space that lead to the correct $Zb\bar b$ couplings while keeping the top and bottom masses to their physical values. We study the predictions for the spectrum of light fermionic resonances and the corrections to $Zt\bar t$ in this scenario.

}

\vfill\eject
\noindent

\section{Introduction}\label{intro}
The main objective of the Large Hadron Collider (LHC) is to discover the real mechanism of electroweak symmetry breaking (EWSB). One of the most interesting solutions for the hierarchy problem is the breaking of the electroweak (EW) symmetry by a strongly interacting sector. Some examples of this approach are Technicolor theories \cite{technicolor} and theories with a composite Higgs boson \cite{old-composite-Higgs}, both scenarios sharing an important difficulty at the theoretical level, the lack of a systematic method to perform precise calculations in strongly coupled theories. Theories with warped extra dimensions have open up new possibilities in this framework, allowing to make perturbative calculations in theories that resemble four dimensional strongly interacting theories \cite{Randall:1999ee,ArkaniHamed:2000ds}.

The main challenge of theories with extra dimensions is to pass the EW precision tests without the generation of a small hierarchy problem, this issue was already present in the original four dimensional proposals \cite{technicolor,old-composite-Higgs}. The EW precision observables have confirmed that the Standard Model (SM) with a light Higgs boson gives an extremely precise description of particle physics \cite{lep-ewwg}. The largest deviation has been observed in the forward-backward asymmetry of the bottom quark $A^b_{FB}$ \cite{Chanowitz:2001bv}, whose measured value deviates by $2.9\sigma$ compared with the SM best fit prediction for the EW observables \cite{lep-ewwg}. On the other hand, $R_b$, the branching fraction of the $Z$ decaying to a pair of bottom quarks, is in very good agreement with the SM prediction. This leads to a challenging puzzle, with potentially new physics effects showing up in $A^b_{FB}$ but giving negligible small corrections for $R_b$. We can translate this situation in terms of the $Zb\bar b$ couplings, obtaining a small pull for $Zb_L\bar b_L$ of order $\sim\delta g^{b_L}\sim 0.003$ and a large pull for $Zb_R\bar b_R$ of order $\sim\delta g^{b_R}\sim 0.02$, compared with the SM best fit prediction \cite{Choudhury:2001hs}~\footnote{There is another possible solution with a large negative shift $\delta g^{b_R}\sim-0.17$. We do not consider that possibility in this work, since it requires very light composite states.}. 

In this work we will show that composite Higgs models can simultaneously solve the $A^b_{FB}$ anomaly and reproduce $R_b$, while giving the correct top and bottom spectrum. We will consider a four dimensional effective theory inspired by extra dimensional theories in a slice of AdS$_5$ \cite{Randall:1999ee}. We will adopt a two sector basis, with the elementary sector reproducing the SM massless spectrum, and the composite sector describing the first lying level of TeV excitations \cite{Contino:2006nn}. The mass eigenstates are mixtures of these sectors, with the lightest states reproducing the SM spectrum. The global symmetry of the composite sector contains the SM gauge symmetry and an extra SU(2)$_R$ global group, with SU(2)$_L\times$SU(2)$_R$ giving rise to the custodial symmetry \cite{custodial}. We will show that the representations under SU(2)$_R$ of the fermionic resonances associated to the third generation of quarks can be chosen to reproduce the observed $A^b_{FB}$ anomaly while preserving $R_b$, with a composite scale of order $\sim 2.7-3.4$ TeV. 

The top mass, arising from the mixing between both sectors, leads to large couplings between the top and the composite sector. Since $t_L$ and $b_L$ are in the same doublet, we expect large corrections for the $b_L$ couplings. As shown in Ref. \cite{Agashe:2006at}, the custodial symmetry can protect the $Zb_L\bar b_L$ coupling from the large corrections induced by the top mixing, resulting in very suppressed modifications of this coupling. Therefore, as noted in Ref. \cite{Bouchart:2008vp}, in the minimal scenario it is in general not possible to obtain the proper shift for $Zb_L\bar b_L$. Moreover, the large correction for $Zb_R\bar b_R$ is in conflict with the bottom mass, since rather strong interactions between the bottom quark and the composite sector are needed for this modification, resulting in a large $m_b$ \footnote{A hierarchical small Yukawa in the composite sector can generate a small bottom mass, but we want to consider a scenario with a natural composite sector, with Yukawa couplings of ${\cal O}(1)$.}. As previously shown in Ref. \cite{Contino:2006qr}, this problem can be solved by introducing two different composite operators to generate the top and bottom masses (see also \cite{Djouadi:2006rk,Bouchart:2008vp}). Therefore, invoking the custodial symmetry, one can choose the SU(2)$_R$ charges of the composite operators generating the top mass to protect $Zb_L\bar b_L$ from the large top corrections and, simultaneously, choose the SU(2)$_R$ charges of the composite operators generating the bottom mass to solve $A^b_{FB}$ while preserving $R_b$.

Similar models in a slice of AdS$_5$ have been proposed in Ref. \cite{Djouadi:2006rk}, with different embeddings for the composite fermions than in the present paper. These models can reproduce the measured $A^b_{FB}$ and $R_b$, but obtained a wrong spectrum for the quarks of the third generation. Another set of models solving the $A^b_{FB}$ anomaly in warped extra dimensions was presented in Ref. \cite{Bouchart:2008vp}, the authors considered different fermionic embeddings than the ones considered in this work. Refs. \cite{Contino:2008hi} and \cite{Mrazek:2009yu} have studied the production and detection of the light fermionic partners of the top at the LHC in this class of models. Refs.~\cite{Carena:2006bn}, \cite{Carena:2007ua} and \cite{Pomarol:2008bh} have investigated the implications of top compositeness, computing the contributions to the Peskin-Takeuchi $T$-parameter \cite{Peskin:1990zt} and determining the allowed degree of top compositeness. They have also focus on the phenomenology calculating the four top production at the LHC. 

The paper is organized as follows: in the next section we describe the model introducing an elementary sector, a composite sector and a Lagrangian mixing both sectors. In section \ref{Zbb} we show the tree level corrections to $Zb\bar b$ in this kind of models. Sections \ref{sym} and \ref{nosym} contain the most important results, there we show different models with and without a symmetry protecting $Zb_L\bar b_L$, discuss the main issues of each approach and show our numerical results. In section \ref{phenomenology} we discuss some important predictions for the phenomenology, like the spectrum of lightest excitations and the corrections to $Zt\bar t$. In section \ref{5d} we briefly describe a five dimensional completion of the effective two sector model. We conclude in section \ref{conclusions} and compare our results with other similar scenarios already present in the literature.

\section{Effective Model}\label{effmodel}
We consider an effective model with elementary fields describing the SM particle content and a composite sector with composite fields (we include the Higgs in the composite sector). The elementary sector has a [SU(3)$_c\times$SU(2)$_L\times$U(1)$_Y$]$^{el}$ gauge symmetry and the composite sector has a [SU(3)$_c\times$SU(2)$_L\times$SU(2)$_R\times$U(1)$_X$]$^{cp}$ global symmetry, with the Higgs transforming as a bidoublet of [SU(2)$_L\times$SU(2)$_R$]$^{cp}$. The elementary fermions have the same quantum numbers as the SM fermions, whereas the composite fermions fulfill complete representations of the composite symmetry. From now on we closely follow the general description of Ref. ~\cite{Contino:2006nn}, see also Refs. \cite{Contino:2008hi,Pomarol:2008bh}. Considering the terms up to dimension four, the Lagrangian can be written as:
\begin{equation}
{\cal L}={\cal L}_{el}+{\cal L}_{comp}+{\cal L}_{mix}
\end{equation} 
with
\begin{eqnarray}
{\cal L}_{el}&=&-\frac{1}{4}F^{el\ 2}_{\mu\nu}+\bar\psi^{el}_L i \Dslash \psi^{el}_L+\bar{\tilde\psi}^{el}_R i \Dslash \tilde\psi^{el}_R \ , \\
{\cal L}_{cp}&=&-\frac{1}{4}F^{cp\ 2}_{\mu\nu}+\frac{m_A^2}{2}A^{cp\ 2}_\mu+\bar\psi^{cp}(i\Dslash^{cp}-m_\psi)\psi^{cp}+\bar{\tilde\psi}^{cp}(i\Dslash^{cp}-\tilde m_{\tilde\psi})\tilde\psi^{cp} \nonumber\\
& & +|D^{cp}_\mu \Sigma|^2-V(\Sigma)-y_{cp}\bar\psi^{cp}\Sigma\tilde\psi^{cp}+{\rm h.c.} \ , \\
{\cal L}_{mix}&=&\frac{m_A^2}{2}\left(-2\frac{g_{el}}{g_{cp}}A^{el}_\mu {\cal P}_AA^{cp}_\mu+\frac{g_{el}^2}{g_{cp}^2}A^{el\ 2}_\mu\right)+\bar\psi^{el}_L \Delta_\psi {\cal P}_\psi \psi^{cp}_R+\bar{\tilde\psi}^{el}_R \tilde\Delta_{\tilde\psi} {\cal P}_{\tilde\psi} \tilde\psi^{cp}_L +{\rm h.c.} \ \label{Lmix},
\end{eqnarray} 
\noindent where the indices $el$ and $cp$ denote elementary and composite fields and couplings, $D$ ($D^{cp}$) is the covariant derivative with respect to $A^{el}$ ($A^{cp}$) and the composite Higgs is $\Sigma=(\tilde H,H)$. A sum over $\psi$ and $\tilde\psi$ is understood, this sum is over the SM Left- and Right-handed fermions, respectively. The elementary fermions are coupled linearly with the composite fermions, and the operators ${\cal P}_{A,\psi,\tilde\psi}$ project the composite bosons and fermions into components with the quantum numbers of the elementary bosons and fermions, with $Y=T^{3R}+T^X$. Once we choose the charges of the composite fermions under [SU(2)$_L\times$SU(2)$_R\times$U(1)$_X$]$^{cp}$ the model is fixed. The parameter space spans over the elementary and composite couplings, the composite masses and the fermionic mixings.

The non-trivial mixings preserve the SM gauge symmetry and give rise to massless gauge bosons and fermions (before EWSB): $A_\mu$, $\psi_L$ and $\tilde\psi_R$, defined by:
\begin{eqnarray}
\begin{bmatrix}A_\mu \\ A_\mu^*\end{bmatrix}=\begin{bmatrix}\cos\theta_A & \sin\theta_A \\ -\sin\theta_A & \cos\theta_A \end{bmatrix} \begin{bmatrix} A^{el}_\mu \\ {\cal P}_A A^{cp}_\mu \end{bmatrix} \ , 
\qquad \tan\theta_A=\frac{g_{el}}{g_{cp}} \ , \label{ASM}\\
\begin{bmatrix}\psi_L \\ \psi_L^*\end{bmatrix}=\begin{bmatrix}\cos\theta_\psi & \sin\theta_\psi \\ -\sin\theta_\psi & \cos\theta_\psi \end{bmatrix} \begin{bmatrix} \psi_L^{el} \\ {\cal P}_\psi \psi_L^{cp} \end{bmatrix} \ , 
\qquad \tan\theta_\psi=\frac{\Delta_\psi}{m_\psi} \ , \label{psiLSM}\\
\begin{bmatrix}\tilde\psi_R \\ \tilde\psi_R^*\end{bmatrix}=\begin{bmatrix}\cos\theta_{\tilde\psi} & \sin\theta_{\tilde\psi} \\ -\sin\theta_{\tilde\psi} & \cos\theta_{\tilde\psi} \end{bmatrix} \begin{bmatrix} \tilde\psi_R^{el} \\ {\cal P}_{\tilde\psi} \tilde\psi_R^{cp} \end{bmatrix} \ , 
\qquad \tan\theta_{\tilde\psi}=\frac{\tilde\Delta_{\tilde\psi}}{\tilde m_{\tilde\psi}} \ .\label{psiRSM} 
\end{eqnarray} 
\noindent The field combinations orthogonal to the massless ones: $A^*_\mu$, $\psi^*$ and $\tilde\psi^*$, have masses:
\begin{equation}\label{M}
M_\phi=\frac{m_\phi}{\cos\theta_\phi}\ , \qquad \phi=A,\psi,\tilde\psi\ .
\end{equation} 
The composite fields that do not mix with the elementary fields (before EWSB) can be written as:
\begin{equation} 
\tilde{\cal P}_\phi \phi^{cp}\ , \qquad \tilde{\cal P}_\phi\equiv 1-{\cal P}_\phi\ , \qquad \phi=A,\psi,\tilde\psi\ ,
\end{equation} 
\noindent and have masses 
\begin{equation}\label{custodianmass}
M_{\tilde{\cal P}\phi}=m_\phi=M_\phi \cos\theta_\phi, 
\end{equation}
where we have considered $m_\phi$ as a function of $M_\phi$. Hereafter we will consider $M_\phi$ as the fundamental parameter. The composite fermions $\tilde{\cal P}_\psi \psi^{cp}$ and $\tilde{\cal P}_{\tilde\psi} \tilde\psi^{cp}$ are usually called custodians.

If we fix the mass of the composite fields $\phi^*$, denoted as $M_\phi$ in Eq.~(\ref{M}), the mass of their SU(2)$_R$ partners $\tilde{\cal P}_\phi \phi^{cp}$ is controlled by the mixing angle $\theta_\phi$. We are interested in the limit $g\ll g_{cp}\ll 4\pi$, thus in general we will have $\sin\theta_A\ll1$, implying that the mass scale of all the composite vector bosons is of order $M_A$, no matter if they are of type ${\cal P}_A A^{cp}$ or $\tilde{\cal P}_A A^{cp}$. However, for the heavy SM fermions as the top, it will be necessary to consider the case $\sin\theta_\psi\sim1$. As previously noted in Ref.~\cite{Pomarol:2008bh}, in this limit, the custodians ${\cal P}_\psi \psi^{cp}$ become very light, and the corresponding massless state $\psi_L$ becomes mostly composite (a similar situation holds for $\psi_R$, exchanging $\psi\to\tilde\psi$).

The Yukawa couplings of the massless fermions $\psi_L$ and $\tilde\psi_R$ are given by:
\begin{equation}\label{yukawa}
y_\psi=y_{cp}\sin\theta_\psi\sin\theta_{\tilde\psi} \ .
\end{equation} 
From the top mass $m_t=y_t v/\sqrt{2}$ we obtain a lower bound on the mixing angles of the top. After EWSB there are new mixings between the SM states and the heavy resonances, as well as mixings between the three generations arising from non-diagonal composite Yukawa couplings. However we do not attempt to make a theory of flavour in this work, for this reason we will consider diagonal Yukawa couplings and neglect the mixing effects.

The mixings of Eq.~(\ref{Lmix}) preserve a SU(3)$_c\times$SU(2)$_L\times$U(1)$_Y$ gauge symmetry. The gauge boson couplings are given by:
\begin{equation}\label{gSM}
g=\frac{g_{el}g_{cp}}{\sqrt{g_{el}^2+g_{cp}^2}} \ ,
\end{equation} 
\noindent where we have omitted a group index for notation simplicity.

After diagonalization of the elementary/composite mixing the massless fermions couple with the composite vector bosons $A^*$ with couplings depending on the mixing angle~\cite{Contino:2006nn}:
\begin{equation}\label{gA*}
{\cal L}\supset - g\ \bar\psi_L \Aslash^*(\cos^2\theta_\psi\tan\theta_A-\sin^2\theta_\psi\cot\theta_A) \psi_L + \{L\leftrightarrow R,\psi\leftrightarrow\tilde\psi\}\ .
\end{equation} 
Note that for $\psi_L$ almost elementary this interaction does not depend on $\theta_\psi$ at leading order, thus we obtain an almost universal coupling for the fermions with small mixing, approximately given by $-g \tan\theta_A$, that is suppressed compared with the gauge coupling $g$ for $g\ll g_{cp}$. On the other hand, if $\psi_L$ is mostly composite, the coupling is large and it is approximately given by $g \cot\theta_A\sim g_{cp}$.

Inspired by the warped extra dimensional description, and for simplicity, we will assume that there is just one common mass scale for all the composite sector, $M_\phi=M$, the only exception being the Higgs, whose mass is determined by the dynamics responsible for $V(\Sigma)$. For similar reasons we will consider that there is a single mixing angle in the gauge sector for all the symmetry groups, thus we will trade 
\begin{equation}
\tan\theta_A\to\tan\theta \ .
\end{equation}

Finally, note that the SM limit can be obtained by taking the composite scale $M\to\infty$, with finite mixing angles $\theta_{A,\psi,\tilde\psi}$. In this limit we recover a theory with massless fermions $\psi_L$ and $\tilde\psi_R$ and gauge bosons $A$ (before EWSB), and finite Yukawa couplings given by Eq.~(\ref{yukawa}).

\section{Corrections to $Zb\bar b$}\label{Zbb}
\begin{figure}
\begin{center}
\begin{picture}(300,110)
        \Photon(-5,50)(44,50){3}{4}
        \CArc(50,50)(6,0,360)
        \Photon(56,50)(105,50){3}{4}
	\Vertex(105,50){3}
        \DashLine(105,50)(89,94){4}
        \CArc(88,100)(6.2,0,360)
        \Line(93,95)(83,104)
        \Line(93,104)(83,95)
        \DashLine(105,50)(121,94){4}
        \CArc(122,100)(6.2,0,360)
        \Line(117,95)(127,104)
        \Line(117,104)(127,95)
        \Photon(105,50)(155,50){3}{4}
	\Vertex(155,50){3}
	\ArrowLine(155,50)(185,95)
	\ArrowLine(185,5)(155,50)
        \CArc(191,95)(6,0,360)
        \CArc(191,5)(6,0,360)
	\ArrowLine(197,95)(247,95)
	\ArrowLine(247,5)(197,5)
        \Text(20,30)[]{$Z^{el}$}
        \Text(75,30)[]{$Z^{cp}$}
        \Text(130,30)[]{$L_3^{cp},R_3^{cp}$}
        \Text(180,60)[]{$\psi^{cp}$}
        \Text(220,75)[]{$\psi^{el}$}
\end{picture}
\caption{Bosonic contribution to $\delta g^\psi$ in the elementary/composite basis, the black dots correspond to composite interactions, the empty dots to elementary/composite insertions and the circled crosses to Higgs vev insertions.}
\label{deltag-gauge}
\end{center}
\end{figure}
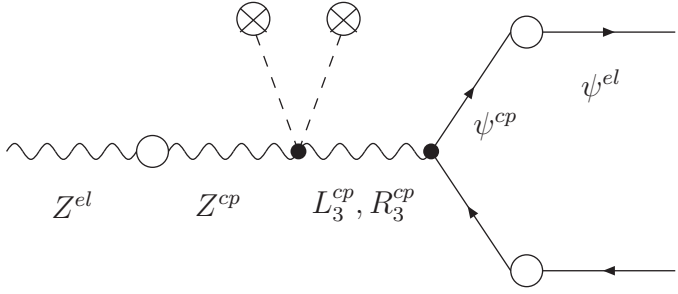

We consider in this section the corrections to the $Z\psi\bar\psi$ couplings in the model described in the previous section. We will compute only the tree level modifications, they are generated by the mixings between the elementary sector and the composite states. In a perturbative expansion, we can consider bosonic and fermionic corrections separately. Although we will consider large mixings for the third generation of fermions, the perturbative approach will provide us useful information to select the fermion embedding leading to the correct phenomenology. We describe first the contributions arising from the mixings between vector bosons after EWSB. We perform a perturbative expansion in insertions of the Higgs vacuum expectation value (vev) and in insertions of the elementary/composite mixing. In Fig.~\ref{deltag-gauge} we show the leading contribution arising from Higgs vev insertions in the bosonic sector.
The elementary/composite insertion in the bosonic leg mixes $Z^{el}$ with $Z^{cp}$, where $Z^{el}$ is the usual combination between $B^{el}$ and $W_3^{el}$, and $Z^{cp}$ is the corresponding combination of composite vector fields (note that defining $B^{cp}$ as the field associated with $T^{3R}+T^X$, $B^{cp}$ can be written as a combination of $R_3^{cp}$ and $X^{cp}$). The Higgs vev adds a new mixing with $L_3^{cp}$ and $R_3^{cp}$. The composite bosons couple with a pair of composite fermions $\psi^{cp}$ that mix with the elementary fermions $\psi^{el}$. We obtain~\cite{Agashe:2006at}
\begin{equation}\label{deltagg}
\delta g^\psi \simeq \frac{g}{c_w} \Delta g_\psi[T^{3R}({\cal P}_\psi\psi^{cp})-T^{3L}({\cal P}_\psi\psi^{cp})]
\end{equation}
with $\Delta g_\psi$ defined by:
\begin{equation}
\Delta g_\psi=\frac{g_{cp}^2}{4}\frac{v^2}{m_A^2}\frac{\Delta_\psi^2}{m_\psi^2} \ .
\end{equation}
There is also a contribution to $\delta g^\psi$ arising from Higgs vev insertions in the fermionic lines. In this case the Higgs vev mixes the different composite fermions $\psi^{cp}$ with equal electric charge, inducing a shift in the $Z\psi\bar \psi$ coupling~\cite{Agashe:2006at}: 
\begin{equation}\label{deltagf}
\delta g^\psi \simeq \frac{g}{c_w} \sum_i |\alpha_{i}|^2[T^{3L}(\psi^{cp}_i)-T^{3L}(\psi^{el})] \ ,
\end{equation}
where $\psi^{cp}_i$ are the composite fermions mixing with $\psi^{el}$ through Higgs vev and elementary/composite insertions, $\alpha_{i}$ is the mixing coefficient between $\psi^{el}$ and $\psi^ {cp}_{i}$. $\alpha_i$ can be obtained by diagonalizing the full mass matrix after EWSB, at perturbative level $\alpha_i\sim y_{cp}v\Delta_\psi/m_\psi m_{\psi_i}$.

\subsection{The $A^b_{FB}$ anomaly and the spectrum of the third generation}
The experimental data on $A_{FB}^b$ and $R_b$ prefer a shift in the $Zb\bar b$ couplings of order $\delta g^{b_L}\sim-0.01 g^{b_L}_{SM}\sim0.003$ and $\delta g^{b_R}\sim0.25 g^{b_R}_{SM}\sim0.02$. To obtain such $\delta g^{b_R}$ we need a large mixing between $\tilde b^{el}_R$ and the composite states, this means a large $\sin\theta_b$, whereas to obtain $\delta g^{b_L}$ we require a smaller mixing between $b^{el}_L$ and the composite states, controlled by $\sin\theta_q$, as can be seen from Eqs.~(\ref{deltagg}) and~(\ref{deltagf}). 
In the simplest model where $q^{el}_L=(t^{el}_L,b^{el}_L)^t$ mixes with just one composite multiplet $q^{cp}$, there is just one parameter controlling the size of the interactions of $t_L$ and $b_L$ with the heavy resonances, $\sin\theta_q$. In particular this parameter controls simultaneously the top and bottom Yukawa, Eq.~(\ref{yukawa}), as well as the corrections to $Zb_L\bar b_L$. As previously discussed, to obtain the large top mass we need a sizable $\sin\theta_q$, such large $\sin\theta_q$ has at least two important consequences: ({\it i}) it gives a large $\delta g^{b_L}$, typically one order of magnitude larger than the desired $\delta g^{b_L}$, unless there is some symmetry protecting $Zb_L\bar b_L$~\cite{Agashe:2006at}, ({\it ii}) since $\sin\theta_b$ is also large, it gives a large Yukawa for the bottom from Eq.~(\ref{yukawa}), leading to a large $m_b$.~\footnote{One can also consider that $y_{cp}^b$ is small to achieve a small Yukawa for the bottom, however in this work we are interested in the limit where the composite couplings satisfy $1\lesssim g_{cp}\lesssim 4\pi$, thus all the small numbers arise from small mixings.} It is possible to invoke a symmetry to protect the $Zb_L\bar b_L$ coupling, but in this case we find that $\delta g^{b_L}$ is much smaller than the value favoured by the phenomenology. Thus, in the simplest case where each elementary field mixes with just one composite field, it is not possible to obtain simultaneously the desired spectrum for the third generation and the proper $Zb\bar b$ couplings in a natural way. 

To relax this tension one can consider a more general situation, where $q^{el}_L$ mixes with two different composite multiplets $q^{1cp}$ and $q^{2cp}$, which can have different charges under SU(2)$_R^{cp}$~\cite{Contino:2006qr,Bouchart:2008vp}. The elementary fields $\tilde t^{el}_R$ and $\tilde b^{el}_R$ mix with just one composite multiplet each, $\tilde u^{cp}$ and $\tilde b^{cp}$ respectively. In this case the Yukawa couplings of the top and bottom sectors are generated by terms involving different fields
\begin{equation}\label{yukawa2}
{\cal L}_{cp}\supset-y_{cp}^t\bar q^{1cp}\Sigma \tilde t^{cp}-y_{cp}^b\bar q^{2cp}\Sigma \tilde b^{cp}+{\rm h.c.} \ . 
\end{equation} 
The top and bottom Yukawa couplings are disentangled and, after diagonalization of the elementary/composite mixings, we get
\begin{equation}\label{yukawa3}
y_t=y_{cp}^t\sin\theta_{q1}\sin\theta_{t} \ , \qquad y_b=y_{cp}^b\sin\theta_{q2}\sin\theta_{b}\ ,
\end{equation} 
with each Yukawa controlled by a different Left-handed mixing angle. The small bottom mass arises as a consequence of a small $\sin\theta_{q2}$.

To implement the previous idea we modify ${\cal L}_{mix}$ and consider two mixing terms for $q_L^{el}$:
\begin{equation}\label{lmixq12}
{\cal L}_{mix}\supset \bar q_L^{el}(\Delta_1 {\cal P}_{q_L}q^{1cp}+\Delta_2 {\cal P}_{q_L}q^{2cp})+\rm{h.c.} \ .
\end{equation} 
Another possibility is to introduce two elementary fields $q_L^{1el}$ and $q_L^{2el}$, both transforming as ${\bf 2}_{1/6}$ under [SU(2)$_L\times$U(1)$_Y$]$^{el}$, and each mixing with the corresponding composite field $q^{1cp}$ and $q^{2cp}$. To get read of one elementary doublet we introduce a Right-handed doublet $q_R^{el}$ with the same quantum numbers as $q_L^{1el}$ and $q_L^{2el}$, together with a mass term marrying $q_R^{el}$ with a linear combination of the Left-handed elementary doublets
\begin{equation}\label{lmassq12}
{\cal L}_{el}\supset m_{el} \ \bar q_R^{el}\ (q_L^{1el} \ \cos\gamma-q_L^{2el} \ \sin\gamma)+\rm{h.c.} \ .
\end{equation} 
The orthogonal combination $q_L^{el}=(q_L^{1el} \ \sin\gamma+q_L^{2el} \ \cos\gamma)$ remains massless up to mixings with the composite sector, this $q_L^{el}$ gives rise to the SM quark doublet after elementary/composite diagonalization. We assume the elementary mass scale to be much larger than the composite scale, of the order of the UV cut-off of the elementary sector.

\section{A symmetry for top contributions to $Zb\bar b$}\label{sym}
To protect the $Zb_L\bar b_L$ coupling from too large corrections arising from the mixings with $q^{1cp}$, we can invoke a $P_{LR}$ symmetry~\cite{Agashe:2006at} (it exchanges $L\leftrightarrow R$), assigning the following charges for ${\cal P}_{b_L}q^{1cp}$, the component of $q^{1cp}$ that mixes with $b^{el}_L$ in Eq.~(\ref{Lmix}):
\begin{equation}\label{ZbLbL}
T^L({\cal P}_{b_L}q^{1cp})=T^R({\cal P}_{b_L}q^{1cp})=1/2 \ , \qquad T^{3L}({\cal P}_{b_L}q^{1cp})=T^{3R}({\cal P}_{b_L}q^{1cp})=-1/2 \ .
\end{equation} 
\noindent Thus we obtain:
\begin{eqnarray}\label{tLsymmetry}
q^{1cp}=({\bf 2},{\bf 2})_{2/3}=
\begin{bmatrix}
 X^{cp'}_1 & U^{cp}_1 \\
 U^{cp'}_1 & D^{cp}_1 
\end{bmatrix} \ ,
\end{eqnarray} 
\noindent where the fields $U_1$ and $D_1$ have electric charges $Q=2/3$ and $Q=-1/3$ respectively, and mix with $q^{el}_L$ in ${\cal L}_{mix}$, the prime fields $U'_1$ and $X'_1$ are the custodians, with charges $Q=2/3$ and $Q=5/3$. In the rest of the paper, to fix the notation for the components of the composite multiplets, we will use ' superindices for the custodial fermions, $D,\ U,\ X$ letters will denote fermions with charges $Q=-1/3,2/3,5/3$, and the subindex will label the composite multiplet to which they belong.

The assignment for $q^{1cp}$ fixes the charges of $t^{cp}$ when we demand the top Yukawa term from Eq.~(\ref{yukawa2}) to be a singlet under the composite symmetries. There are two possibilities for $t^{cp}$:
\begin{eqnarray}
{\rm[}T1{\rm]}& & t^{cp}=({\bf 1},{\bf 3})_{2/3}+({\bf 3},{\bf 1})_{2/3}=
\begin{bmatrix} X^{cp'}_t & U^{cp}_t & D^{cp'}_t \end{bmatrix}+
\begin{bmatrix} X^{cp''}_t \\ U^{cp'}_t \\ D^{cp''}_t \end{bmatrix}\ ,\label{tembedding1} \\
{\rm[}T2{\rm]}& & t^{cp}=({\bf 1},{\bf 1})_{2/3}=U^{cp}_t  \ .\label{tembedding2}
\end{eqnarray}
We will denote with $T1$ and $T2$ the embeddings for the top sector defined by (\ref{tLsymmetry},\ref{tembedding1}) and (\ref{tLsymmetry},\ref{tembedding2}) respectively.
With this assignments for $q^{1cp}$ and $t^{cp}$, the contributions from the top sector to $\delta g^{b_L}$ are very small, although there remains an irreducible contribution arising from the symmetry breaking by ${\cal L}_{mix}$~\cite{Casagrande:2010si}.
The leading corrections to $Zb\bar b$ arise from the mixings with the bottom sector, $q^{2cp}$ and $b^{cp}$. In section~\ref{nosym} we will discuss how to obtain the proper corrections to $Zb\bar b$ without relying in symmetries to cancel the corrections from the top sector.

\subsection{Bottom embedding}\label{bottomembedding}
We discuss now the embeddings of the bottom sector able to reproduce the desired shifts in the $Zb\bar b$ couplings. From Eq.~(\ref{deltagg}) we obtain a positive $\delta g^{b_L}$ from the gauge sector if
\begin{equation}\label{condbLg}
T^{3R}({\cal P}_{b_L}q^{2cp})>-1/2 \ ,
\end{equation} 
where $T^{3R}({\cal P}_{b_L}q^{2cp})$ is the SU(2)$_R$ charge of the composite fermion in $q^{2cp}$ that mixes with $b^{el}_L$. We also obtain a positive $\delta g^{b_R}$ from the gauge sector if
\begin{equation}\label{condbRg}
T^{3R}({\cal P}_{b_R}b^{cp})>0 \ .
\end{equation} 
Since $Y=T^{3R}+T^X$, choosing $T^{3R}$ we fix also the U(1)$_X$ charge. Eqs.~(\ref{condbLg}) and~(\ref{condbRg}), together with the condition of a gauge invariant bottom Yukawa operator, lead us to the following possible embeddings for the bottom sector
\begin{eqnarray}
{\rm[}B1{\rm]}& q^{2cp}=({\bf 2},{\bf 3})_{-5/6}=
\begin{bmatrix}
 U^{cp}_2 & D^{cp'}_2 & V^{cp''}_2 \\
 D^{cp}_2 & V^{cp'}_2 & S^{cp'}_2
\end{bmatrix} \ , 
&  b^{cp}=({\bf 1},{\bf 2})_{-5/6}=\begin{bmatrix} D^{cp}_b & V^{cp'}_b\end{bmatrix} \ , \label{bembedding1}\\
{\rm[}B2{\rm]}& q^{2cp}=({\bf 2},{\bf 4})_{-4/3}=
\begin{bmatrix}
 U^{cp}_2 & D^{cp'}_2 & V^{cp''}_2 & S^{cp''}_2 \\
 D^{cp}_2 & V^{cp'}_2 & S^{cp'}_2 & T^{cp'}_2
\end{bmatrix}, 
& b^{cp}=({\bf 1},{\bf 3})_{-4/3}=\begin{bmatrix} D^{cp}_b & V^{cp'}_b & S^{cp'}_b\end{bmatrix} \ \ \ \label{bembedding2}
\end{eqnarray}
where $V,S$ and $T$ are exotic fermions with $Q=-4/3,-7/3$ and $-10/3$, respectively. Is is possible to consider larger representations also. We have considered only the cases where $q^{2cp}$ and $b^{cp}$ are respectively a doublet and a singlet of SU(2)$_L^{cp}$, otherwise ${\cal L}_{mix}$ breaks the SU(2)$_L$ gauge symmetry of the SM. 

As previously mentioned, at perturbative level in vev insertions, there are also fermionic corrections to $Zb\bar b$ arising from the vev mixings between the fermions, Eq.~(\ref{deltagf}). The analysis is similar for both embeddings in Eqs.~(\ref{bembedding1}) and (\ref{bembedding2}), $b_R$ mixes at leading order in vev insertions with the state $|{\cal P}_{b_L}q^{2cp}\rangle\equiv|D_2^{cp}\rangle$, and it also mixes with $|D_2^{cp'}\rangle$, that is the associated down custodian. Since these states have opposite SU(2)$_L$ charges, their contribution to $\delta g^{b_R}$ is partially cancelled, as can be checked using Eq.~(\ref{deltagf}) (they do not cancel exactly because their Yukawa couplings can be different, depending on the corresponding Clebsch-Gordan coefficients, and because the custodians can be lighter than the other resonances, due to the SU(2)$_R$ breaking by ${\cal L}_{mix}$, as discussed in section~\ref{effmodel}). At leading order in vev insertions $b_L$ mixes with the state $|{\cal P}_{b_R}b^{cp}\rangle\equiv|D_b^{cp}\rangle$, inducing a positive $\delta g^{b_L}$.

As a summary, we present in table~\ref{tdeltagsym} the perturbative contributions to $\delta g^{b_L}$ and $\delta g^{b_R}$, at leading order in vev insertions, for the different embeddings $[T1],\ [T2],\ [B1]$ and $[B2]$. We have included also the fermionic contribution to $\delta g^{b_L}$ arising from the top sector, associated to the heavy resonances $D_t^{cp'}$ and $D_t^{cp''}$ from $t^{cp}$. The custodial symmetry ensures an effective cancellation between these contributions. Note that, parametrically, the shifts from the gauge contributions in embeddings $[B1]$ and $[B2]$ follow the pattern $\delta g^{b_L}_{[B1]}/\delta g^{b_L}_{[B2]}=3/4$ and $\delta g^{b_R}_{[B1]}/\delta g^{b_R}_{[B2]}=1/2$, meaning that it is easier to achieve a large correction in case $[B2]$.
\begin{table}[h] 
\begin{center} 
\begin{tabular}{|c|c|c|} 
\hline
embedding & $\delta g^{b_L}/(g/c_w)$ & $\delta g^{b_R}/(g/c_w)$ \\[.1in] \hline
$[T1]$ & $\frac{1}{2}(\alpha_{D_t'}^2-\alpha_{D_t''}^2)$ & - \\[.1in]\hline
$[T2]$ & - & - \\[.1in]\hline
$[B1]$ & $\frac{3}{2}\Delta g_{q2}+\frac{1}{2}\alpha_{D_b}^2$ & 
$\frac{1}{2}\Delta g_{b}+\frac{1}{2}(\alpha_{D_2'}^2-\alpha_{D_2}^2)$ \\[.1in]\hline
$[B2]$ & $2\Delta g_{q2}+\frac{1}{2}\alpha_{D_b}^2$ & 
$\Delta g_{b}+\frac{1}{2}(\alpha_{D_2'}^2-\alpha_{D_2}^2)$ \\[.1in] \hline
\end{tabular} 
\end{center} 
\caption{$\delta g^{b_L}$ and $\delta g^{b_R}$ at leading order expanding in powers of vev insertions. The results correspond to the top and bottom embeddings of Eqs.~(\ref{tembedding1},\ref{tembedding2},\ref{bembedding1}) and (\ref{bembedding2}).}
\label{tdeltagsym}
\end{table}

\subsection{Full diagonalization}
We have given perturbative arguments to show that it is possible to obtain the correct spectrum and $Zb\bar b$ couplings. The perturbative expansion in Higgs vev insertions is a good approximation for the gauge sector, but in general this is not true for the fermionic mixings since in this case $\sin\theta_\psi$ can be $\sim1$. For this reason it is better to make a numerical diagonalization of the mass matrices. In this section we will scan the parameter space and show the configurations that reproduce the observables by performing a full numeric diagonalization in the gauge and fermionic sectors. As mentioned in sec. \ref{effmodel}, we will not consider mixings between the three generations in our approach, that analysis is beyond the scope of this work. However we consider important being able to obtain the proper results in this simplified approach, without relying in potential new contributions from the full approach that could improve the situation.

After EWSB all the states with the same electric charge are mixed. To obtain the couplings between the physical states we have to diagonalize the full mass matrices of the gauge bosons and fermions. These matrices include contributions from the composite sector, the elementary/composite mixing and the contributions from the Higgs vev, that depend on the embeddings. The lightest state in the down and up fermionic sectors correspond to the physical $b$- and $t$-quarks respectively. The photon decouples due to the remaining electromagnetic gauge symmetry and the lightest massive neutral vector state corresponds to the physical $Z$-boson. After rotation from the flavour basis to the physical mass basis in the interaction terms, we obtain the interactions between the physical states.

Let us consider the parameters of the model, since we have chosen a single composite scale $M$ and a common bosonic mixing $\tan\theta_A$, there are 5 parameters in the bosonic sector: $M,\ g_{cp}^c,\ g_{cp}^L,\ g_{cp}^R$ and $g_{cp}^X$. We choose three of them to reproduce the SM gauge couplings, Eq.~(\ref{gSM}), and we choose $g_{cp}^L=g_{cp}^R$ to implement the $P_{LR}$ symmetry in the composite sector. We have made a random scan over the fermionic parameters, fixing $m_b=4.2$ GeV and $m_t=175$ GeV, we have considered composite Yukawas $|y_{cp}^{t,b}|<2\pi$ and mixings $0<\sin\theta_{q1,q2,t,b}<1$. 

\begin{figure}[ht] \centering
\psfrag{dgL}[t]{$10^2\times\delta g^{b_L}$}
\psfrag{dGR}[bl][tc]{$10^2\times\delta g^{b_R}$}
\psfrag{0.0}[t]{\footnotesize0}
\psfrag{0.1}[t]{\footnotesize0.1}
\psfrag{0.2}[t]{\footnotesize0.2}
\psfrag{0.3}[t]{\footnotesize0.3}
\psfrag{0.4}[t]{\footnotesize0.4}
\psfrag{0.5}[t]{\footnotesize0.5}
\psfrag{0.6}[t]{\footnotesize0.6}
\psfrag{0.8}[t]{\footnotesize0.8}
\psfrag{1.5}[t]{\footnotesize1.5}
\psfrag{1.2}[t]{\footnotesize1.2}
\psfrag{1.4}[t]{\footnotesize1.4}
\psfrag{1.6}[t]{\footnotesize1.6}
\psfrag{1.8}[t]{\footnotesize1.8}
\psfrag{2.5}[t]{\footnotesize2.5}
\psfrag{3.5}[t]{\footnotesize3.5}
\psfrag{1.0}[t]{\footnotesize1}
\psfrag{2.0}[t]{\footnotesize2}
\psfrag{3.0}[t]{\footnotesize3}
\includegraphics[width=.45\textwidth]{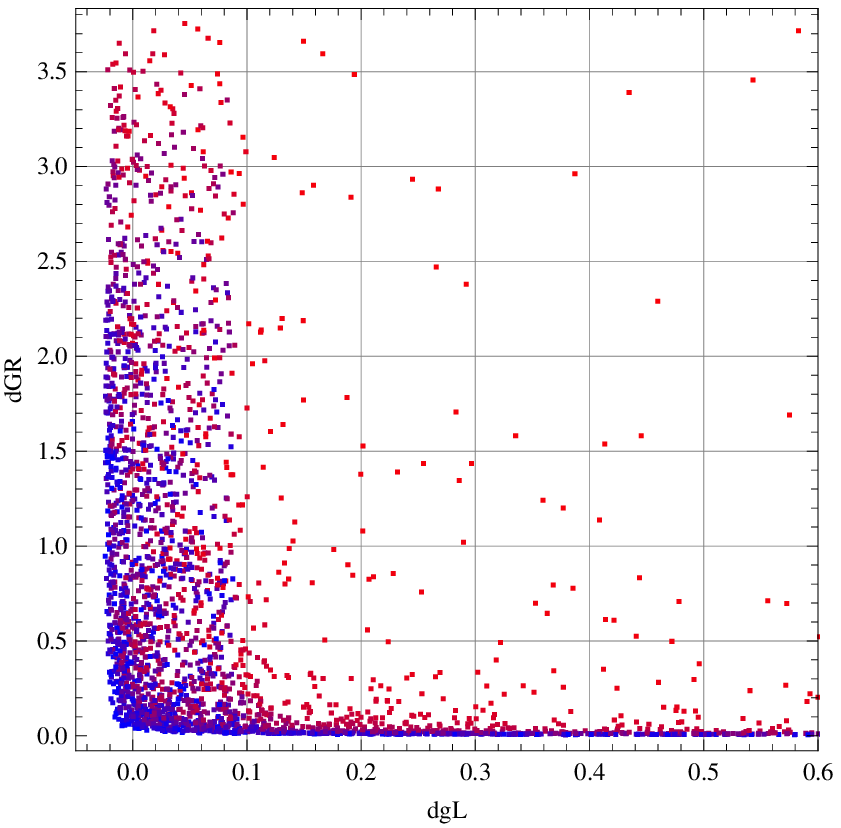}
\hskip1cm
\includegraphics[width=.45\textwidth]{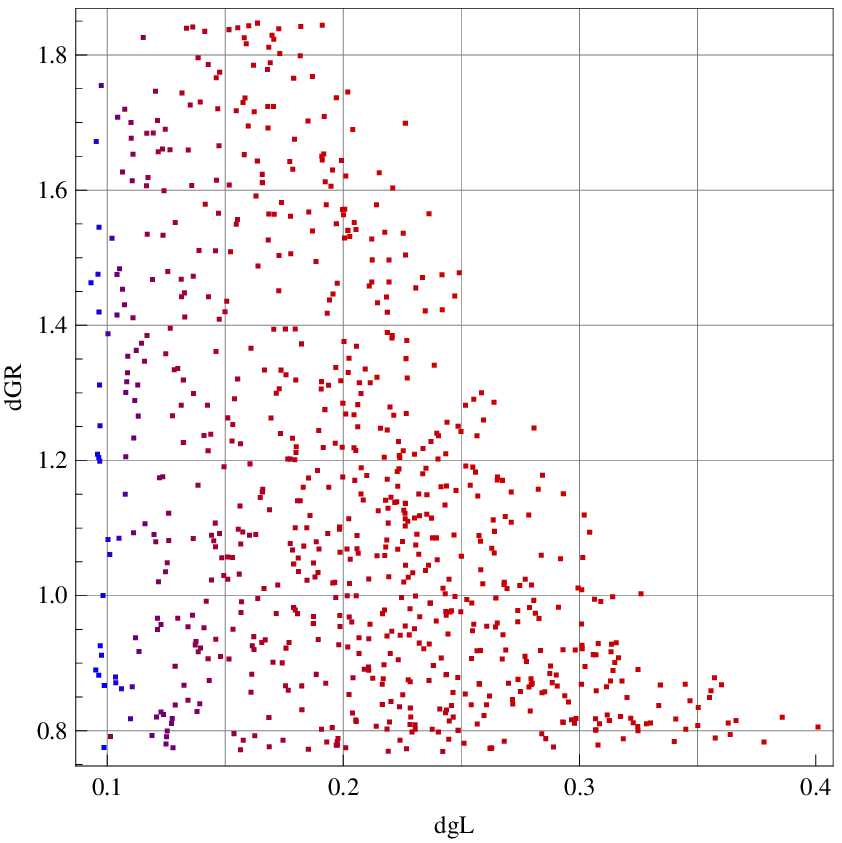}
\vskip0.5cm
\caption{Plan $\delta g^{b_L}/\delta g^{b_R}$ for model $[T2+B2]$, with full numeric diagonalization. We have fixed  $m_t=175$ GeV and $m_b=4.2$ GeV. On the left we show our results for the scan specified in the text, $|y_{cp}^{t,b}|<2\pi$ and mixings $0<\sin\theta_{q1,q2,t,b}<1$. The colors codify the size of $y_{cp}^b$, from red corresponding to $y_{cp}^b=0.2$ to blue corresponding to $y_{cp}^b=2\pi$, with intermediate colors interpolating between those values. On the right we show our results for the scan over a smaller region of the parameter space, $0.45\lesssim \sin\theta_b\lesssim0.70$ and $0.25\lesssim \sin\theta_{q2}\lesssim 0.40$. The colors codify the size of $y_{cp}^b$, from red corresponding to $y_{cp}^b=0.3$ to blue corresponding to $y_{cp}^b=1.6$, with intermediate colors interpolating between those values.}
\label{figsymdeltag}
\end{figure}

We discuss first our results for the model $[T2+B2]$, with $M=2.7$ TeV and $g_{el}/g_{cp}=1/8$, this ratio corresponds to an SU(2)$_L^{cp}$ coupling $g^{cp}_L\simeq 5.3$. We show our results in Fig.~\ref{figsymdeltag}, on the left we have plotted the shifts $\delta g^{b_L}$ and $\delta g^{b_R}$, fixing the third generation quark masses to their physical values and varying randomly the fermionic mixings and the Yukawa couplings as explained previously. The first thing we want to note is that the order of magnitude of $\delta g^{b_L}$ and $\delta g^{b_R}$ are the correct ones. Most of the points are localized in the region of small $\delta g^{b_L}$, showing a tension with the bottom mass. There also some points with small $\delta g^{b_R}$ and larger $\delta g^{b_L}$. The colors codify the size of the composite bottom Yukawa, $y_{cp}^b$, with red and blue corresponding to $y_{cp}^b=0.2$ and $y_{cp}^b=2\pi$, respectively, and intermediate colors interpolating between these extreme values.   

The value of $\delta g^{b_R}$ is driven by $\sin\theta_b$, whereas $\delta g^{b_L}$ is mostly determined by $\sin\theta_{q2}$. Demanding $0.0075\lesssim \delta g^{b_R}\lesssim 0.0225$ and $0.001\lesssim \delta g^{b_L}\lesssim 0.004$ we obtain $0.45\lesssim \sin\theta_b\lesssim0.70$ and $0.25\lesssim \sin\theta_{q2}\lesssim 0.40$. On the right of Fig.~\ref{figsymdeltag} we show our results for the fermionic mixings restricted to these intervals. The colors codify the size of the composite bottom Yukawa, $y_{cp}^b$, with red and blue corresponding to $y_{cp}^b=0.3$ and $y_{cp}^b=1.6$, respectively, and intermediate colors interpolating between these extreme values. For $\sin\theta_b$ and $\sin\theta_{q2}$ in these ranges, we find that a small Yukawa is preferred in the bottom sector: $0.3\lesssim y_{cp}^b\lesssim 1$, as can be seen in the plot by the small number of blue points and the large number of almost red ones. It is possible to find solutions with $y_{cp}^b\sim 1.5$, but the region of the parameter space with $y_{cp}^b\gtrsim 1$ is smaller than the region with $y_{cp}^b\lesssim 1$. Thus, even in the model with two different composite fields associated to the elementary Left-handed doublet $q_L^{el}$, we observe  some tension because $\delta g^{b_L}$ prefers $\sin\theta_{q2}$ not too small, whereas $m_b$ prefers a smaller $\sin\theta_{q2}$ (or a small composite Yukawa if we keep $\sin\theta_{q2}$ large enough to obtain the right $\delta g^{b_L}$, as we have done in this plot). 

Summarizing, in the models with a symmetry to cancel the top contributions to $g^{b_L}$, if we adjust the bottom mixings to reproduce the shifts in the $Zb\bar b$ couplings, the composite bottom Yukawa $y_{cp}^b$ has to be somewhat smaller than the composite top Yukawa $y^t_{cp}$, with $y^b_{cp}/y^t_{cp}\sim 0.1-0.7$. Despite this tension, we remark that in these models it is possible to find solutions with all the composite couplings of ${\cal O}(1)$. 

If we decrease $\tan\theta_A$ the corrections for the $Zb\bar b$ couplings are smaller, as can be seen in Eqs.~(\ref{deltagg}) and~(\ref{deltagf}). For $M=2.7$ TeV and $\tan\theta_A=1/5$ we obtain results similar to Fig.~\ref{figsymdeltag}, but in this case we need larger mixings $0.6\lesssim \sin\theta_b\lesssim1$ and $0.35\lesssim \sin\theta_{q2}\lesssim 0.65$ to achieve the right corrections for the couplings. The bottom mass imposes as a constraint that the larger the mixings the smaller the composite Yukawa couplings $y_{cp}^b$, thus in this case we obtain $0.1\lesssim y_{cp}^b\lesssim 0.5$, introducing some tuning in the composite sector. Note that $y_{cp}^b\ll g_{cp},y_{cp}^t$, since the top mass gives us a lower bound $y_{cp}^t\gtrsim 2$ and we are considering $g_{cp}\sim 3$. A similar effect is observed if we fix $\tan\theta_A=1/8$ but increase $M$. As an example, for $M=3.4$ TeV the same range of mixings and Yukawas as in the previous case are needed to obtain the right couplings and spectrum, introducing again some tuning in the composite sector. Another possibility is to decrease $\tan\theta_A$ and $M$ simultaneously, in this case there is a partial compensation between the effects of a smaller elementary/composite mixing in the gauge sector and a lower composite scale. However for smaller $M$ we obtain larger corrections to other EW precision observables, like the $S$ and $T$ parameter, as well as flavour observables, that prefer a larger $M$.

We have also investigated embedding $[T1]$ for the $t_R$ partner in the composite sector. This case is very similar to embedding $[T2]$ if we impose $P_{LR}$ for the Yukawa couplings of the $t_R$ partners transforming as $({\bf 1},{\bf 3})_{2/3}$ and $({\bf 3},{\bf 1})_{2/3}$, leading to $y_{cp}^{({\bf 1},{\bf 3})}= y_{cp}^{({\bf 3},{\bf 1})}$. In this case, basically all the analysis we have done for the previous model is valid for embedding $[T1]$ also, and for this reason we do not show the corresponding results in the $g^{b_L}/g^{b_R}$ plan. However, in embedding $[T1]$ we can obtain a larger $\delta g^{b_L}$ without modification of the other important observables if we allow a small violation of the $P_{LR}$ symmetry, {\it i.e.:} $y_{cp}^{({\bf 1},{\bf 3})}\neq y_{cp}^{({\bf 3},{\bf 1})}$. Thus the small bottom mass is not in conflict with $\delta g^{b_L}$, since we can obtain the proper $m_b$ by adjusting $\sin\theta_{q2}$ and the right $\delta g^{b_L}$ by considering a small $(y_{cp}^{({\bf 1},{\bf 3})}- y_{cp}^{({\bf 3},{\bf 1})})\neq0$, in such a way that the different observables are controlled by different parameters, and $y^b_{cp}$ can be larger than in the completely symmetric case. Note that this approach differs from the previous one, where just the bottom sector was responsible for achieving the shifts in the $Zb\bar b$ couplings. In the present approach the bottom sector produces $\delta g^{b_R}$ and the top sector is responsible for $\delta g^{b_L}$. If we want $\delta g^{b_L}$ to be $\sim 0.003$, we need the difference between the top Yukawas to be not larger than $\sim10\%$, introducing again some tuning in the composite sector.

We have also analysed the embedding $[B1]$ for $q^{2cp}$ and $b^{cp}$. As mentioned at the end of sec. \ref{bottomembedding}, the corrections to $Zb\bar b$ are parametrically smaller than in embedding $[B2]$. Then larger fermionic mixings and a small $y_{cp}^b$ are required in the $P_{LR}$ symmetric approach. For this reason the situation is somewhat worse than in the case of embedding $[B2]$. 

\subsection{Radiative corrections to $Zb\bar b_L$}
We have computed the corrections to $Zb\bar b$ at tree level, however the large degree of compositeness of the top sector and the light custodians can induce important one-loop corrections. Although radiative corrections are beyond the scope of this work, it is crucial to estimate the size of the most important loop contributions in our model. At one loop, the size of the corrections to $Zb_L\bar b_L$ can be estimated by considering the gaugeless limit and computing the coupling $H^3b_L\bar b_L$, with $H^3$ the neutral Goldstone eaten by the $Z$. The corrections from the new sector are generated by the composite fermions running in the loop~\cite{Oliver:2002up}. As shown in Ref.~\cite{Carena:2007ua}, the dominant contributions arise from the top sector, specially from the fermions $U^{cp'}_1$ and $U^{cp}_1$ in $q^{1cp}$ and $U^{cp}_t$ in $t^{cp}$. Although the symmetry protecting $Zb_L\bar b_L$ is broken by ${\cal L}_{mix}$, there is a partial cancellation between $U^{cp'}_1$ and $U^{cp}_1$, therefore $U^{cp}_t$ dominates in general. We have checked that these corrections give $\delta g^{b_L}\sim 3\times10^{-4}$ for the region of the parameter space solving $A_{FB}^b$ at tree level, with just a few points giving $\delta g^{b_L}\sim (1-3)\times10^{-3}$. We have also estimated the size of the corrections arising from a neutral composite vector resonance running in the loop. These contributions are suppressed by the smaller mixings and Yukawa in the bottom sector, leading to $\delta g^{b_L}\sim 10^{-5}$. Although a full one-loop calculation is needed, our estimates suggest that the tree level corrections dominate. 

\section{No symmetry for $Zb\bar b$}\label{nosym}
We will show in this section that it is possible to obtain the proper corrections to $Zb\bar b$ without relying in symmetry arguments. However, as we will show, in general in this case one has to invoke accidental cancellations or introduce tuning in the composite sector to achieve the proper $\delta g^{b_L}$.

We find two possibilities to obtain a small correction from the top sector: ({\it i}) there is a partial cancellation between the gauge and fermionic corrections to $\delta g^{b_L}$, ({\it ii}) there is no fermionic contribution at leading order because there is no resonance with $Q=-1/3$ in $t^{cp}$, thus at leading order there are just bosonic contributions.

\subsection{Partial cancellation of top sector correction to $Zb\bar b$}
The first option is realized when the gauge and fermionic contributions to $\delta g^{b_L}$ have opposite signs. An example is given by the following embedding for the top sector
\begin{eqnarray}\label{tembeddingcancel}
[T3] & q^{1cp}=({\bf 2},{\bf 3})_{7/6}=
\begin{bmatrix}
 Y^{cp'}_1 & X^{cp''}_1 & U^{cp}_1 \\
 X^{cp'}_1 & U^{cp'}_1 & D^{cp}_1
\end{bmatrix} \ , \nonumber \\
 &\ \ \ t^{cp}=({\bf 1},{\bf 4})_{7/6}=\begin{bmatrix} Y^{cp'}_t & X^{cp'}_t & U^{cp}_t & D^{cp'}_t\end{bmatrix} \ ,
\end{eqnarray}
\noindent with $Y'$ being exotic fermions with $Q=8/3$.

At leading order in vev and elementary/composite insertions the gauge contribution to $\delta g^{b_L}$ is negative and the fermionic one is positive, as can be seen in table \ref{tdeltagnosym}. Besides these corrections there are corrections arising from the bottom sector, as already shown in table \ref{tdeltagsym} for embeddings $[B1]$ and $[B2]$ of $q_2^{cp}$ and $b^{cp}$.
\begin{table}[h] 
\begin{center} 
\begin{tabular}{|c|c|} 
\hline
top-embedding & $\delta g^{b_L}/(g/c_w)$ \\[.1in] \hline
$[T3]$ & $-\frac{1}{2}\Delta g_{q1}+\frac{1}{2}\alpha_{D_t'}^2$ \\[.1in]\hline
$[T4]$ & $-\Delta g_{q1}+\frac{1}{2}\alpha_{D_t'}^2$ \\[.1in]\hline
$[T5]$ & $-\frac{1}{2}\Delta g_{q1}$ \\[.1in]\hline
\end{tabular} 
\end{center} 
\caption{$\delta g^{b_L}$ from the top sector at leading order expanding in powers of vev insertions. The results correspond to the top embeddings of Eqs.~(\ref{tembeddingcancel}), (\ref{tembeddingcancel2}) and (\ref{tembeddingnof}).}
\label{tdeltagnosym}
\end{table}

\begin{figure}[ht] \centering
\psfrag{dgL}[t]{$10^2\times\delta g^{b_L}$}
\psfrag{dGR}[bl][tc]{$10^2\times\delta g^{b_R}$}
\psfrag{sQ1}[t]{$\sin \theta_{q1}$}
\psfrag{dgL1}[bl][tc]{$10^2\times\delta g^{b_L}$}
\psfrag{0.0}[t]{\footnotesize0}
\psfrag{0.1}[t]{\footnotesize0.1}
\psfrag{0.2}[t]{\footnotesize0.2}
\psfrag{0.3}[t]{\footnotesize0.3}
\psfrag{0.4}[t]{\footnotesize0.4}
\psfrag{0.5}[t]{\footnotesize0.5}
\psfrag{0.6}[t]{\footnotesize0.6}
\psfrag{0.7}[t]{\footnotesize0.7}
\psfrag{0.8}[t]{\footnotesize0.8}
\psfrag{0.9}[t]{\footnotesize0.9}
\psfrag{1.5}[t]{\footnotesize1.5}
\psfrag{1.2}[t]{\footnotesize1.2}
\psfrag{1.4}[t]{\footnotesize1.4}
\psfrag{1.6}[t]{\footnotesize1.6}
\psfrag{1.8}[t]{\footnotesize1.8}
\psfrag{2.5}[t]{\footnotesize2.5}
\psfrag{3.5}[t]{\footnotesize3.5}
\psfrag{1.0}[t]{\footnotesize1}
\psfrag{2.0}[t]{\footnotesize2}
\psfrag{3.0}[t]{\footnotesize3}
\psfrag{3.0}[t]{\footnotesize3}
\psfrag{0}[t]{\footnotesize0}\psfrag{1}[t]{\footnotesize1}\psfrag{2}[t]{\footnotesize2}\psfrag{3}[t]{\footnotesize3}\psfrag{4}[t]{\footnotesize4}\psfrag{5}[t]{\footnotesize5}\psfrag{6}[t]{\footnotesize6}
\includegraphics[width=.45\textwidth]{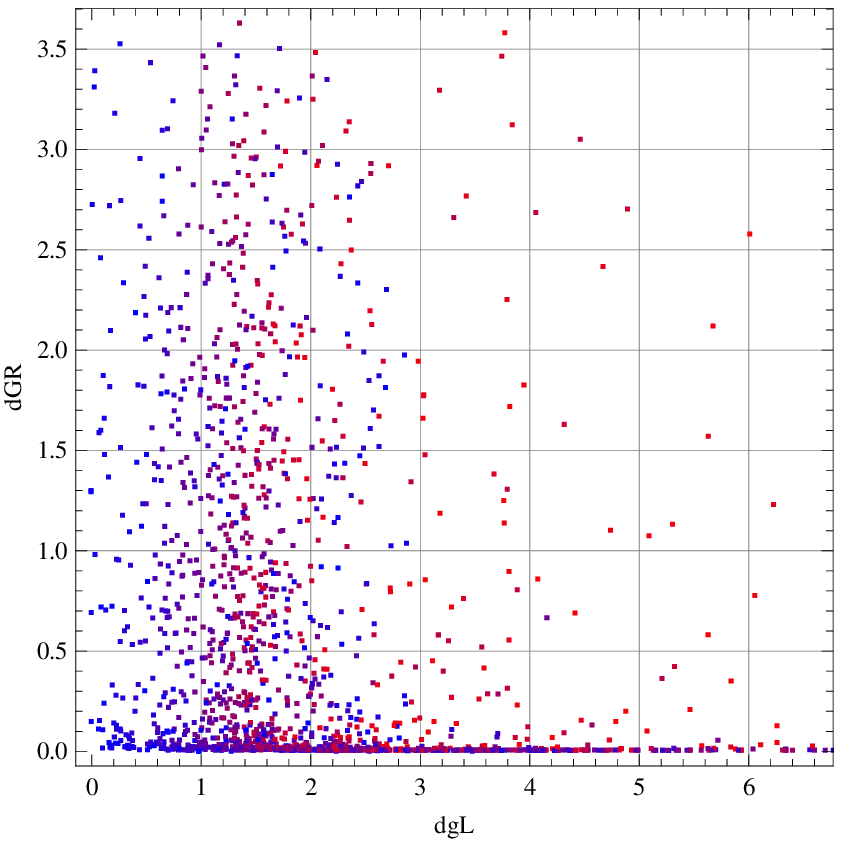}
\hskip1cm
\includegraphics[width=.45\textwidth]{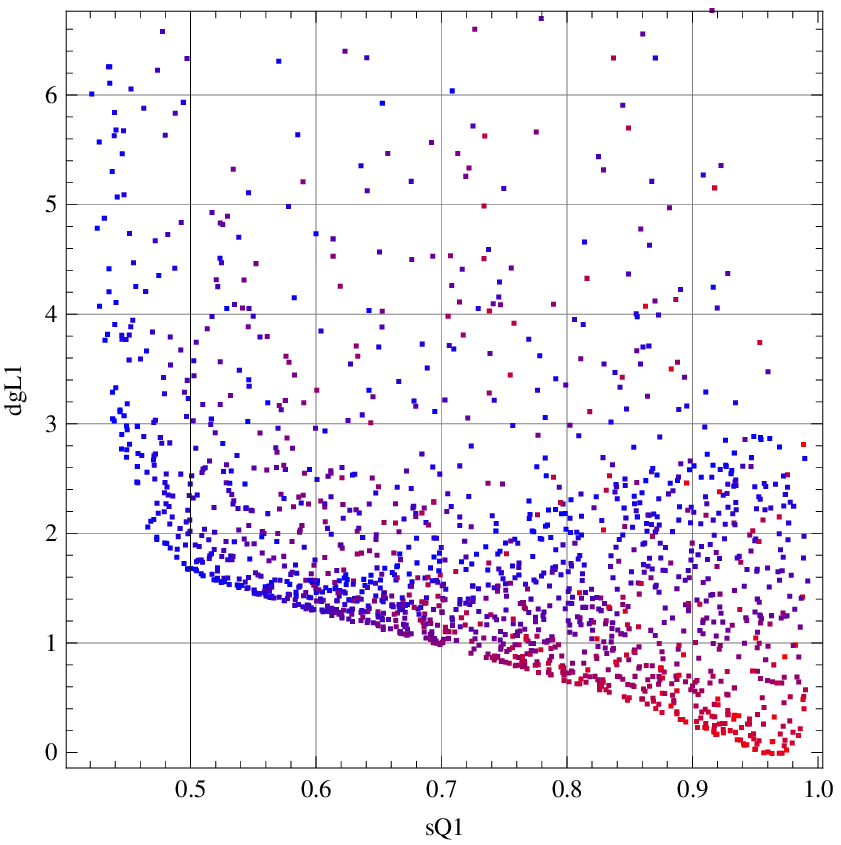}
\vskip0.5cm
\caption{Plan $\delta g^{b_L}/\delta g^{b_R}$ for model $[T3+B2]$, with full numeric diagonalization. We have fixed  $m_t=175$ GeV and $m_b=4.2$ GeV. On the left we show our results for the scan specified in the text, $|y_{cp}^{t,b}|<2\pi$ and mixings $0<\sin\theta_{q1,q2,t,b}<1$. The colors codify the size of $\sin\theta_{q1}$, from red corresponding to $\sin\theta_{q1}\simeq0.41$ to blue corresponding to $\sin\theta_{q1}\simeq0.99$, with intermediate colors interpolating between those values. On the right we show the dependence of $\delta g^{b_L}$ with $\sin\theta_{q1}$. The colors codify the size of $y_{cp}^t$, with red and blue corresponding to the minimum and maximum couplings, $y_{cp}^t\simeq2.5$ and $y_{cp}^t=2\pi$, and intermediate colors interpolating between those values.}
\label{figcanceldeltag}
\end{figure}

The first thing we want to note is that in this model it is also possible to obtain the correct spectrum and $Zb\bar b$ corrections. However it is not natural to obtain a small $\delta g^{b_L}$, unless there is a partial cancellation between the contributions arising from the top sector. Except for some small regions of the parameter space, in general the gauge and fermionic contributions to $\delta g^{b_L}$ are one order of magnitude larger than the desired shift. Therefore, it requires some tuning to obtain an effective cancellation in the absence of symmetries, and the natural shift results one order of magnitude larger than the needed one. We show our results in Fig.~\ref{figcanceldeltag}, where we have made a scan similar to the one of Fig~\ref{figsymdeltag}, but now with the embeddings $[T3+B2]$ for the top and bottom composite sectors, respectively. We have made a random scan over the fermionic parameters, fixing $m_b=4.2$ GeV and $m_t=175$ GeV, with $|y_{cp}^{t,b}|<2\pi$, $0<\sin\theta_{q1,q2,t,b}<1$, $M=2.7$ TeV and $g_{el}/g_{cp}=1/8$. On the left we have plotted $\delta g^{b_L}$ and $\delta g^{b_R}$. We can see that the $Zb\bar b$ corrections are spread over a much larger range than in the symmetric case, in fact most of the points are in the region of $\delta g^{b_L}\sim 10^{-2}$, that is one order of magnitude larger than in the previous models. The fermionic contribution to $\delta g^{b_L}$ from $t^{cp}$ is positive and large, so the negative gauge contribution from $q^{1cp}$ has to be also large in order to obtain an approximate cancellation. This situation can be better visualized in the right plot of Fig.~\ref{figcanceldeltag}, where we show $\delta g^{b_L}$ in terms of $\sin\theta_{q1}$, for the same scan as in the left plot. The colors codify the size of $y_{cp}^t$, with red and blue corresponding to the minimum and maximum couplings, $y_{cp}^t\simeq2.5$ and $y_{cp}^t=2\pi$, and intermediate colors interpolating between those values. The dispersion is due to the random variation of the other parameters. A small $\delta g^{b_L}$ can be obtained only in the limit of $\sin\theta_{q1}\simeq1$, and with the smallest $y_{cp}^t$ compatible with the top mass, this assignments maximize the negative contribution and minimize the positive one, allowing an effective cancellation.

A similar situation holds for a model with embeddings $[T3+B1]$, but in this case it is more difficult to obtain the proper $Zb\bar b$, since the corrections from the bottom sector are parametrically smaller, as in the case with $P_{LR}$ symmetry.

Another example where an accidental cancellation can be present is
\begin{eqnarray}\label{tembeddingcancel2}
[T4] & q^{1cp}=({\bf 2},{\bf 4})_{5/3}=
\begin{bmatrix}
 Z^{cp'}_1 & Y^{cp'}_1 & X^{cp''}_1 & U^{cp}_1 \\
 Y^{cp''}_1 & X^{cp'}_1 & U^{cp'}_1 & D^{cp}_1
\end{bmatrix}, \nonumber \\
& t^{cp}=({\bf 1},{\bf 5})_{5/3}=\begin{bmatrix} Z^{cp'}_t & Y^{cp'}_t & X^{cp'}_t & U^{cp}_t & D^{cp'}_t\end{bmatrix} \ ,
\end{eqnarray}
\noindent with $Z$ being exotic fermions with $Q=11/3$.

As shown in table \ref{tdeltagnosym}, the gauge contribution to $\delta g^{b_L}$ arising from $q^{1cp}$ is larger than in case $[T3]$. Thus it is possible to obtain a cancellation with the fermionic contribution from $t^{cp}$ for lower values of $\sin\theta_{q1}$. However we find the same problem as before, the absence of a symmetry protecting $\delta g^{b_L}$ implies that only in a very small region of the parameter space the cancellation is effective.

\subsection{No fermionic correction to $Zb\bar b$ from the top sector}
In this section we present an embedding such that there is no down-type quark in $t^{cp}$, thus, at leading order in vev insertions, the only contribution from the top sector to $\delta g^{b_L}$ arises from gauge mixing. Since we want a positive $\delta g^{b_L}$, it would be preferable to obtain a positive contribution to $\delta g^{b_L}$ from the top sector. However we did not find any embedding satisfying that condition and without a down-type quark in $t^{cp}$. Let us consider the following embedding for the top sector
\begin{eqnarray}\label{tembeddingnof}
[T5] & q^{1cp}=({\bf 2},{\bf 3})_{7/6}=
\begin{bmatrix}
 Y^{cp'}_1 & X^{cp''}_1 & U^{cp}_1 \\
 X^{cp'}_1 & U^{cp'}_1 & D^{cp}_1
\end{bmatrix} \ , 
& \ t^{cp}=({\bf 1},{\bf 2})_{7/6}=\begin{bmatrix} X^{cp'}_t & U^{cp}_t \end{bmatrix} \ .
\end{eqnarray}
In table \ref{tdeltagnosym} we show the contribution to $\delta g^{b_L}$ arising from $[T5]$. The top mass requires large mixings with the composite sector, thus the negative gauge contribution to $\delta g^{b_L}$ arising from the top sector is large. To obtain $\delta g^{b_L}\sim 0.003$ we need a sizable positive contribution from the bottom sector, that can be realized if the mixings with $q^{2cp}$ are also large. As usual, we need some tuning in order to obtain a small $\delta g^{b_L}$ emerging from the cancellation between the negative top correction and the positive bottom one. To reduce the top contribution we can consider the minimum $\sin\theta_{q1}$ and the maximum $y_{cp}^t$ compatible with the top mass and a perturbative expansion. On the other hand, $\delta g^{b_R}$ is mostly driven by the mixings with $b^{cp}$, requiring a sizable $\sin\theta_b$. Summarizing, to obtain the proper corrections to $Zb\bar b$ the mixings with $q^{2cp}$ and $b^{cp}$ have to be large. Since these mixings also determine the bottom mass, the right $m_b$ can be obtained only if $y_{cp}^b$ is small enough, introducing a small hierarchy in the composite sector: $y_{cp}^b\ll y_{cp}^t$.

We have analysed this model by making a full numerical diagonalization. We have performed a random scan over the fermionic parameters, fixing $m_b=4.2$ GeV and $m_t=175$ GeV, with $|y_{cp}^{t,b}|<2\pi$, $0<\sin\theta_{q1,q2,t,b}<1$, $M=2.7$ TeV and $g_{el}/g_{cp}=1/8$. We find that in general $\delta g^{b_L}$ turns out to be negative and rather large, $\delta g^{b_L}\sim-0.005$. We find solutions for the shifts of the $Zb\bar b$ couplings and the spectrum in the following region of the parameter space: $0.4\lesssim \sin\theta_b\lesssim0.8$, $0.3\lesssim \sin\theta_{q2}\lesssim 0.55$ and $0.1\lesssim y_{cp}^b\lesssim 0.2$ for the bottom sector and $0.5\lesssim \sin\theta_{t}\lesssim1$, $0.3\lesssim \sin\theta_{q1}\lesssim 0.7$ and $4\lesssim y_{cp}^b\lesssim 2\pi$ for the top sector, however we stress that in this region there also many points that do not give the proper $\delta g^{b_{L,R}}$. In fact the density of points with wrong $\delta g^{b_{L,R}}$ is much larger than the density of points with the correct $\delta g^{b_{L,R}}$, as expected since some fine tuning is needed to obtain the small $\delta g^{b_L}$ out of the accidental cancellation between the larger contributions from the different sectors.

\section{Phenomenology}\label{phenomenology}
In this section we briefly mention some distinctive phenomenological signals associated with the solutions for the $A^b_{FB}$ anomaly. We discuss first some properties of the spectrum of resonances. 

\subsection{Spectrum of light custodians}
\begin{figure}[ht] \centering
\psfrag{dgL}[t]{$10^2\times\delta g^{b_L}$}
\psfrag{m}[bl][tc]{$M_\psi$[TeV]}
\psfrag{dgL1}[bl][tc]{$10^2\times\delta g^{b_L}$}
\psfrag{0.0}[t]{\footnotesize0}
\psfrag{0.1}[t]{\footnotesize0.1}
\psfrag{0.2}[t]{\footnotesize0.2}
\psfrag{0.3}[t]{\footnotesize0.3}
\psfrag{0.4}[t]{\footnotesize0.4}
\psfrag{0.5}[t]{\footnotesize0.5}
\psfrag{0.6}[t]{\footnotesize0.6}
\psfrag{0.7}[t]{\footnotesize0.7}
\psfrag{0.8}[t]{\footnotesize0.8}
\psfrag{0.9}[t]{\footnotesize0.9}
\psfrag{1.5}[t]{\footnotesize1.5}
\psfrag{1.2}[t]{\footnotesize1.2}
\psfrag{1.4}[t]{\footnotesize1.4}
\psfrag{1.6}[t]{\footnotesize1.6}
\psfrag{1.8}[t]{\footnotesize1.8}
\psfrag{2.5}[t]{\footnotesize2.5}
\psfrag{3.5}[t]{\footnotesize3.5}
\psfrag{1.0}[t]{\footnotesize1}
\psfrag{2.0}[t]{\footnotesize2}
\psfrag{3.0}[t]{\footnotesize3}
\psfrag{3.0}[t]{\footnotesize3}
\psfrag{0}[t]{\footnotesize0}\psfrag{1}[t]{\footnotesize1}\psfrag{2}[t]{\footnotesize2}\psfrag{3}[t]{\footnotesize3}\psfrag{4}[t]{\footnotesize4}\psfrag{5}[t]{\footnotesize5}\psfrag{6}[t]{\footnotesize6}
\includegraphics[width=.45\textwidth]{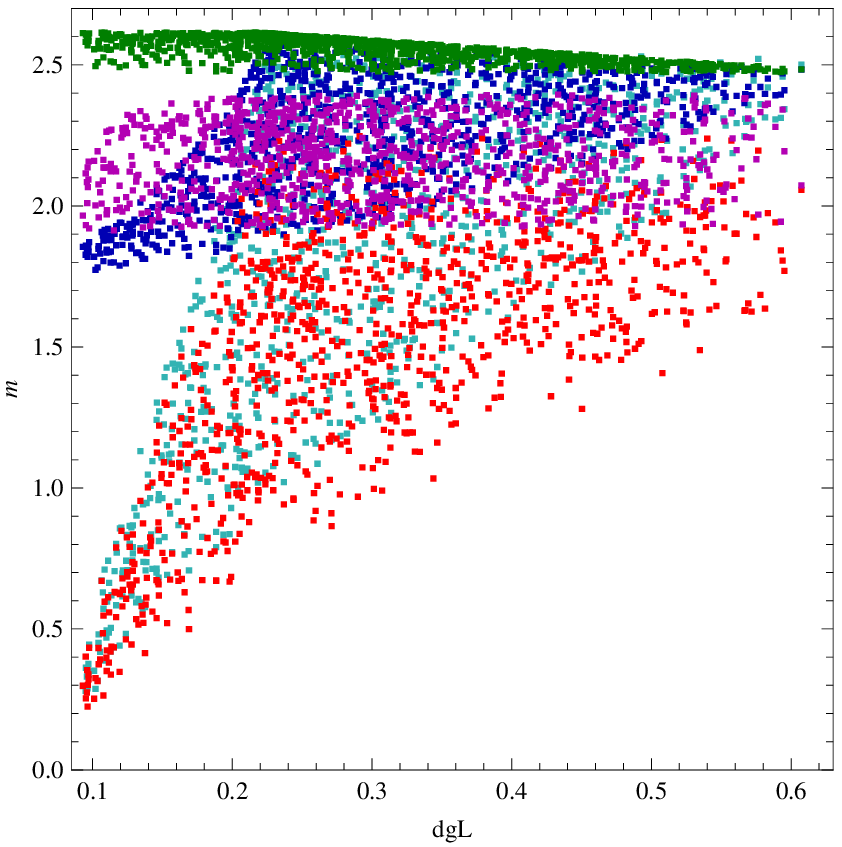}
\hskip1cm
\includegraphics[width=.45\textwidth]{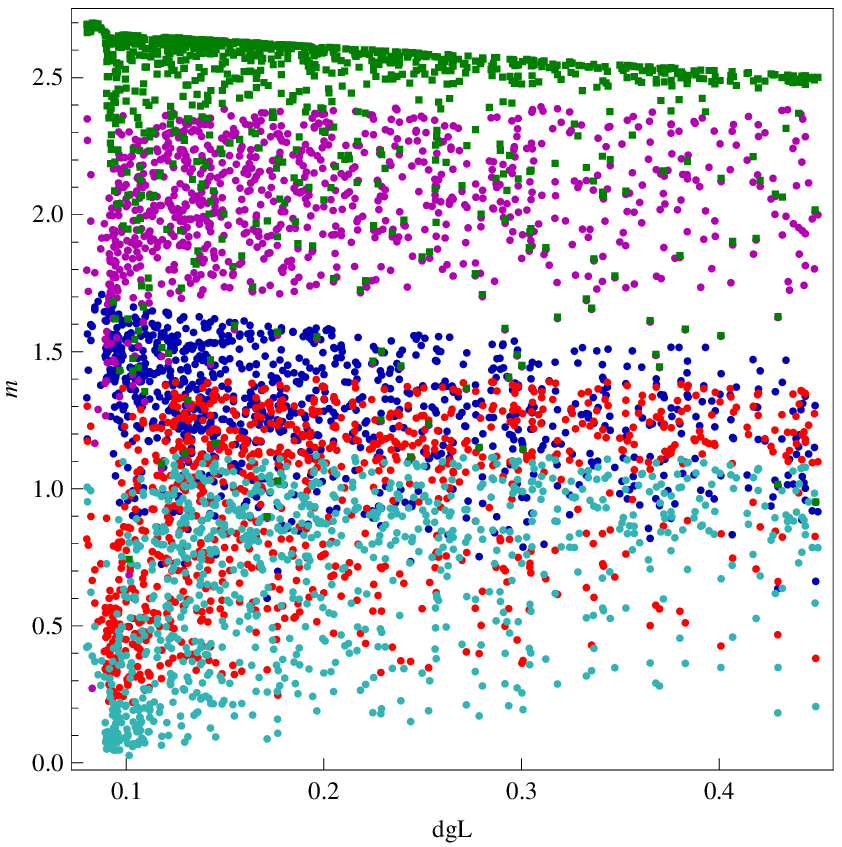}
\vskip0.5cm
\caption{Spectrum of the lightest fermionic resonances in terms of $\delta g^{b_L}$. On the left (right) we show our results for model $[T2+B2]$ ($[T1+B2]$), with full numeric diagonalization. We have fixed  $m_t=175$ GeV, $m_b=4.2$ GeV and $0.0075\lesssim\delta g^{b_R}\lesssim 0.0225$, with $\tan\theta=1/8$ and the composite scale $M=2.7$ TeV. The colors distinguish fermionic resonances with different charges: red for $Q=2/3$, blue for $Q=-1/3$, light blue for $Q=5/3$, violet for $Q=-4/3,-7/3$ and green for $Q=-10/3$.}
\label{figKK}
\end{figure}

In all the models that we have considered, at least one of the top chiralities has a large degree of compositeness, needed to obtain the large top Yukawa~(\ref{yukawa}). In fact this is a common feature of most of the models where the top mass arises from the mixings with a composite sector \cite{Carena:2006bn,Contino:2006qr,Pomarol:2008bh}. We have also shown that, to protect the $Zb\bar b$ interactions from large corrections arising from the top sector, the top composite partners must furnish non trivial representations of SU(2)$_R$, giving rise to custodian fermions. As can be seen in Eq.~(\ref{custodianmass}), the mass of the custodians is suppressed by $\cos\theta_\psi$ compared with the composite scale $M$, thus the suppression becomes important for large mixings $\theta_\psi\sim\pi/2$, leading to light resonances associated to the top sector. We analyse first model $[T2+B2]$, with custodians associated to $t_L$ only, and no custodians for $t_R$. Before EW symmetry breaking the custodians in the top sector can be identified with $U^{cp'}_1$ and $X^{cp'}_1$ in Eq.~(\ref{tLsymmetry}), and the custodian mass is given by Eq.~(\ref{custodianmass}). After EWSB, all the fermions with a given electric charge are mixed by the Yukawa couplings, introducing corrections to Eq.~(\ref{custodianmass}). For the region of the parameter space that gives the correct spectrum and $Zb\bar b$ corrections, we find light resonances with $Q=2/3$ and $Q=5/3$, arising mostly from $U^{cp'}_1$ and $X^{cp'}_1$, their masses being almost degenerate, with a dispersion of order $10\%$ due to different Clebsch-Gordan coefficients in the Yukawa couplings. For $\tan\theta=1/8$ and $M=2.7$ TeV these light resonances have masses in the range $\sim 0.3-2.2$ TeV, depending on the value of $\theta_{q1}$. There are custodians in the bottom sector also. Since $b^{cp}$ is a triplet of SU(2)$_R$, there are two custodians $V_b^{cp'}$ and $S_b^{cp'}$, with $Q=-4/3$ and $Q=-7/3$ respectively, associated to $b_R$. In order to obtain the large $\delta g^{b_R}$ we need moderate mixings between $b_R$ and $b^{cp}$, obtaining a mild mass suppression for these custodians. After EW symmetry breaking we find almost degenerate exotic resonances with $Q=-4/3,-7/3$, and for $\tan\theta=1/8$ and $M=2.7$ TeV their masses are in the range $1.9-2.4$ TeV. Finally, the small $\delta g^{b_L}$ demand small mixings between $b_L$ and $q^{2cp}$, thus the masses of the custodians arising from $q^{2cp}$ are not suppressed. After EWSB and for $\tan\theta=1/8$ and $M=2.7$ TeV we find masses in the range $2.4-2.6$ TeV for these fermions. A summary of these results is presented in Fig. \ref{figKK}, where we show the spectrum of the lightest resonances and $\delta g^{b_L}$, using different colors to distinguish resonances with different charges. 

Since $\delta g^{b_R}$ is mostly determined by $\sin\theta_b$, fixing $\delta g^{b_R}\sim 0.02$ we obtain $\sin\theta_b\sim 0.7$. The mass of the custodians arising from $b^{cp}$ is given by $M_{\tilde P b_R}\simeq M \cos\theta_b$, up to corrections arising from the Higgs vev. Therefore, fixing $\delta g^{b_R}\sim 0.02$ we fix the mass of the custodians associated to $b^{cp}$, with charges $-4/3$ and $-7/3$, to be of order $1.9$ TeV in this case. We have checked this estimate by performing a full numeric diagonalization, obtaining masses in the range $1.9-2.0$ TeV, the dispersion due to the variation of the other mixings and Yukawa couplings. The spectrum of the other resonances is not correlated with $\delta g^{b_R}$.

If we consider a smaller $g_{cp}$, with $\tan\theta=1/5$ for example, the custodians arising from the bottom sector become lighter than in the case $\tan\theta=1/8$ when we demand the proper $Zb\bar b$ corrections. The reason is that for smaller $\tan\theta$, larger mixings are needed to correct $Zb\bar b$, obtaining a larger suppression for the custodian masses. In this case, for $M=2.7$ TeV, the custodians arising from $q^{1cp}$ are again almost degenerate with masses within $0.3-2.4$ TeV,  whereas the custodians arising from $q^{2cp}$ are in the range $2.1-2.6$ TeV and the custodians arising from $b^{cp}$ can be as light as $0.2-2.1$ TeV (a region of the parameter space can thus be excluded by direct search limits).

The analysis of the fermionic spectrum is very similar for model $[T1+B2]$. We briefly discuss some issues where the spectrum shows differences with respect to model $[T2+B2]$. In this case $t^{cp}$ is embedded in a larger representation of SU(2)$_R$, Eq. (\ref{tembedding1}), introducing new custodians: $D^{cp'}_t,D^{cp''}_t$ with charge $Q=-1/3$, $U^{cp'}_t$ with $Q=2/3$ and two exotic quarks $X^{cp'}_t$ and $X^{cp''}_t$ with $Q=5/3$. Therefore, since there are $X$- and $U$-type custodians associated to both chiralities of the top, we always find light resonances with charges $Q=5/3$ and $Q=2/3$, the exotic fermions being somewhat lighter due to different Clebsch-Gordan coefficients in the Yukawa couplings. For $\tan\theta=1/8$ and $M=2.7$ TeV we obtain the lightest $X$-type quarks in the range $0.02-1.1$ TeV and the lightest $U$-type quarks in the range $0.3-1.3$ TeV, the region of the parameter space with resonances lighter than $\sim 300$ GeV is already excluded by direct search. Therefore, this model generically predicts exotic and $t$-type quarks with masses below $\sim1$ TeV.
We also find light resonances with $Q=-1/3$, with masses $\sim 0.5-1.7$ TeV, arising from the $D$-type custodians in $t^{cp}$. Another difference is that we find solutions with larger mixings in the bottom sector: $0.5\lesssim\sin\theta_b\lesssim 0.8$ and $0.3\lesssim\sin\theta_{q2}\lesssim 0.9$. Although not all the points in this region give the correct $\delta g^{b_{L,R}}$, there are combinations that are successful, giving rise to lighter exotic $V$-, $S$- and $T$-type resonances, with charges $Q=-4/3,-7/3$ and $-10/3$ respectively. For $\tan\theta=1/8$ and $M=2.7$ TeV the lightest $V$- and $S$-resonances are in the range $0.7-2.4$ TeV and are almost degenerate, the $T$-resonance is in the range $0.75-2.6$ TeV. Our results are shown in Fig. \ref{figKK}, where we have plotted the spectrum of the lightest resonances against $\delta g^{b_L}$. In this case we again obtain a correlation between $\delta g^{b_R}$ and the mass of the custodians arising from $b^{cp}$, obtaining exotic fermions with masses in the range $1.70-1.95$ TeV for $\delta g^{b_R}\sim 0.02$. For this model we have found some solutions with lower masses, up to $\sim 0.5$ TeV, that correspond to larger values of $\sin\theta_b$ or $\sin\theta_{q2}$, and with $\delta g^{b_R}\sim 0.02$. However the number of solutions with very light bottom-custodians is much lower than the number of solutions with masses of order $1.70-1.95$ TeV.

The models obtained considering the other combinations of top and bottom embeddings have similar properties to these cases, namely the presence of rather light custodians, including exotic fermions.

\subsection{Corrections to $Zt\bar t$}
Another important aspect for the phenomenology are the corrections in the $Zt\bar t$ and $Wt\bar b$ interactions. As already mentioned in Ref. \cite{Agashe:2006at}, if we choose a symmetry for $Zb_L\bar b_L$, it is not possible to protect simultaneously the other bottom and top couplings. However, if $T^{3L}({\cal P}_{t_R} t^{cp})=T^{3R}({\cal P}_{t_R} t^{cp})=0$, there is a $P_C$ symmetry that can protect $Zt_R\bar t_R$ \cite{Agashe:2006at}. The $P_C$ symmetry is realized for embeddings $[T1]$ and $[T2]$, in these cases $Zt_L\bar t_L$ and $Wb_L\bar t_L$ are not protected (similar results were obtained in Ref. \cite{Pomarol:2008bh}).

Using the results of sec. \ref{Zbb}, at leading order in powers of vev insertions we obtain the corrections to $Zt\bar t$ shown in table \ref{tdeltagtsym}. We can see explicitly the cancellation between the fermionic contributions to $Zt_R\bar t_R$ for embeddings $[T1]$ and $[T2]$, however due to the SU(2)$_R$ breaking by ${\cal L}_{mix}$ this cancellation is not exact. The bottom sector also modifies $Zt_L\bar t_L$, since $t_L$ arises from the mixings between $t^{el}_L$, ${\cal P}_{t_L}q^{1cp}$ and ${\cal P}_{t_L}q^{2cp}$, with $q^{2cp}$ within the so called bottom sector. Its contributions are also shown in table \ref{tdeltagtsym}.
\begin{table}[h] 
\begin{center} 
\begin{tabular}{|c|c|c|} 
\hline
embedding & $\delta g^{t_L}/(g/c_w)$ & $\delta g^{t_R}/(g/c_w)$ \\[.1in] \hline
$[T1]$ & $-\Delta g_{q1}-\frac{1}{2}(\alpha_{U_t}^2+\alpha_{U_t'}^2)$ & $\frac{1}{2}(\alpha_{U_1}^2-\alpha_{U_1'}^2)$ \\[.1in]\hline
$[T2]$ & $-\Delta g_{q1}-\frac{1}{2}\alpha_{U_t}^2$ & $\frac{1}{2}(\alpha_{U_1}^2-\alpha_{U_1'}^2)$ \\[.1in]\hline
$[T3]$,\ $[T5]$ & $-\frac{3}{2}\Delta g_{q1}-\frac{1}{2}\alpha_{U_t}^2$ & $\frac{1}{2}(\alpha_{U_1}^2-\alpha_{U_1'}^2)-\frac{1}{2}\Delta g_{t}$ \\[.1in]\hline
$[T4]$ & $-2\Delta g_{q1}-\frac{1}{2}\alpha_{U_t}^2$ & $\frac{1}{2}(\alpha_{U_1}^2-\alpha_{U_1'}^2)-\frac{1}{2}\Delta g_{t}$ \\[.1in]\hline
$[B1]$ & $\frac{1}{2}\Delta g_{q2}$ & - \\[.1in]\hline
$[B2]$ & $\Delta g_{q2}$ & - \\[.1in]\hline
\end{tabular} 
\end{center} 
\caption{$\delta g^{t_L}$ and $\delta g^{t_R}$ at leading order expanding in powers of vev insertions. The results correspond to the top and bottom embeddings considered in the previous sections.}
\label{tdeltagtsym}
\end{table}

\noindent We have computed the corrections to $Zt\bar t$ performing a full diagonalization of all the mixings. We present here the numerical results obtained for the models giving the better solutions for $Zb\bar b$, namely the models defined by $[T1+B2]$ and $[T2+B2]$. The other cases are qualitatively similar, and do not offer new insights. Restricted to the region of the parameter space that gives the right corrections to $Zb\bar b$, $[T1+B2]$ gives $\delta g^{t_L}\sim-0.02$, although we find solutions where $\delta g^{t_L}$ can be as large as $\sim -0.1$. For the right-handed top we obtain smaller corrections as expected from the custodial symmetry, $\delta g^{t_R}\sim -0.001$, for some special regions of the parameter space we find solutions with larger negative shifts $\delta g^{t_R}\sim -0.01$ and also solutions with positive shifts $\delta g^{t_R}\sim +0.001$. Embedding $[T2+B2]$ gives similar results for $\delta g^{t_R}$, for $Zt_L\bar t_L$ we find $\delta g^{t_L}\sim -0.015$, although $\delta g^{t_L}\sim -0.05$ and $\delta g^{t_L}\sim +0.001$ are also possible. Note that this numerical results are consistent with the perturbative estimate of table \ref{tdeltagtsym}.

\subsection{LHC phenomenology}
A distinctive signature of our composite solution to $A^b_{FB}$ is the presence of light fermionic resonances with exotic charges. As shown in the previous sections, the lightest resonances are the custodians associated with the top sector. The pair and single production of these custodians, as well as their detection, has been studied in Refs.~\cite{Contino:2008hi},~\cite{Mrazek:2009yu} and \cite{AguilarSaavedra:2009es}. As already discussed, the top-custodians are needed to keep small the top contribution to $Zb_L\bar b_L$. However, the fermionic resonances that drive the proper corrections to $Z b\bar b$, are the custodians associated to the bottom sector, $q^{2cp}$ and $b^{cp}$. The production and detection of these states at the LHC would provide strong evidence for our model. For the embedding $[B2]$, that better reproduces the experimental results for $R_b$ and $A^b_{FB}$, there is a custodian with $Q=-1/3$, as well as exotic custodians with charges $Q=-4/3,-7/3$ and $-10/3$. Although the bottom-custodians are somewhat heavy, compared with the top-ones, the sizable mixing needed to obtain $\delta g^{b_R}\sim 0.02$ induces a suppression in the exotic custodians arising from $b^{cp}$. These states have charges $Q=-4/3$ and $-7/3$, are almost degenerate and, for a composite scale $M=2.7$ TeV, have masses in the range $1.9-2.4$ TeV, although we have found some regions of the parameter space with masses as small as 0.5 TeV. In the following paragraphs we will briefly discuss the production and detection of the resonances $q^*_{-4/3}$ and $q^*_{-7/3}$. For model $[T1+B2]$ there is also the possibility of having a light exotic custodian with charge $Q=-10/3$, that could lead to a striking signal with three $W$'s plus a $b$-jet for each resonance. However, since in general the mass of this custodian is not suppressed, we will not discus its production here. 

The new states $q^*_{-4/3}$ and $q^*_{-7/3}$ could be produced in pairs at the LHC by QCD interactions: $q\bar q, g g\to q^*_{-4/3}\bar q^*_{-4/3},q^*_{-7/3}\bar q^*_{-7/3}$, the cross section strongly depending on the custodian mass. Once created, $q^*_{-4/3}$ will predominantly decay to $W^-$ plus $b_R$, since the mixings with the lighter quarks will be suppressed due to their light masses. The creation of $q^*_{-4/3}\bar q^*_{-4/3}$ will therefore lead to the final state $W^-bW^+\bar b$, with very energetic $b$-jets~\cite{AguilarSaavedra:2009es}. This final state is the same as in the decay of a pair of fermions with charge $2/3$, like $t\bar t$ and heavier resonances. Both processes could be distinguished by reconstructing the charge of the intermediate resonance, for example allowing the $W^+$ to decay leptonically and demanding four jets, two of them identified as $b$-jets. The mass of $q^*_{-4/3}$ will be equal to the invariant mass of the $b$-jet plus the $W^-$-jets. Note that this reconstruction requires the measurement of the charge of the $b$-jets. For the pair creation of $q^*_{-7/3}$, its decay will be dominated by $q^*_{-7/3}\to W^-q^*_{-4/3}\to W^-W^-b$, with a virtual $q^*_{-4/3}$. Thus pair creation of $q^*_{-7/3}$ states will lead to a final state with four $W$'s plus two $b$-jets: $W^-W^-bW^+W^+\bar b$. Notice that the same final state can arise, for example, from the decay of a resonance associated with the top sector with charge $5/3$. Again both processes could be distinguished by reconstructing the charge of the intermediate resonance.

Pair production of $q^*_{-4/3}$ and $q^*_{-7/3}$ resonances will be highly suppressed at the LHC for a mass scale $M_{q^*}\gtrsim 2$ TeV, in this case single production should dominate, since it falls off much more slowly with increasing mass. $q^*_{-4/3}$ can be produced by $bW$ fusion, with the $W$ being radiated by one proton and the $b$ coming from the other proton: $qb\to q'q^*_{-4/3}$. Ref.~\cite{Wb} has shown that $bW$ fusion is an efficient process to singly produce heavy resonances in $pp$ and $p\bar p$ collisions. Single production of $q^*_{-4/3}$ will lead to a final state with $W^-$ plus $b$, and a spectator light quark jet. 
On the other hand, we expect single production of $q^*_{-7/3}$ to be suppressed, since it involves a virtual $q^*_{-4/3}$ resonance.
A careful analysis is required to obtain precise information about these processes and the possibility of reconstructing the resonances, since measuring the $b$-jet charge is difficult.

\section{5D description}\label{5d}
There is a dual picture of the elementary/composite model we have described in the previous sections, namely a 5 dimensional theory compactified in a warped space with boundaries. The study of extra dimensional theories is usually performed by decomposing the 5D fields in an infinite series of 4D fields, the Kaluza-Klein modes. The advantage of this decomposition is that the Kaluza-Klein modes correspond to the 4D mass eigenstate basis. The lightest eigenstates describe the SM fields and there is a tower of heavier excited states corresponding to the rest of the Kaluza-Klein modes. There is also another useful description of a 5D theory with boundaries called the holographic description, where the bulk fields are separated from the boundary fields and they are treated as different degrees of freedom. If the bulk fields are weakly coupled with the boundary fields, it is possible to make a perturbative expansion in terms of the boundary/bulk couplings to study the theory. This actually happens for theories in AdS$_5$, or more generally in spaces that can be approximated by AdS$_5$ near the ultraviolet boundary. Depending on which physical properties one wants to study, either the Kaluza-Klein or the boundary/bulk description offer a simpler and deeper understanding. At low energies, one can consider another simplification, instead of working with the full tower of Kaluza-Klein states, one can truncate the series retaining only the lowest lying excitations. In the boundary/bulk description this would correspond to retain the first Kaluza-Klein level of the bulk degrees freedom only. For processes involving energy $E$, the truncated effective description obtained in this way have errors of order $(E/M_2)^2$, with $M_2$ the scale of the second Kaluza-Klein level, thus for a large family of observables, truncation can be a good approximation. Although this approximation can be systematically improved by including higher excitations, and eventually by summing the effects of the full tower of states, we hope that the simplified approach can capture some of the most important features of the corrections to the $Zb\bar b$ interaction. In this case we obtain an effective 4D theory with two sectors, the first one corresponding to the boundary fields, and the second sector arising from the lightest Kaluza-Klein level of the bulk fields. This framework corresponds to the the approach we have considered in the previous sections, with the lightest Kaluza-Klein level identified as the resonances of the composite sector. 

The truncation we have performed can also be thought as a discretization or deconstruction of the fifth dimension, with only two sites. Increasing the number of sites the extra dimension emerges.

As suggested by the AdS/CFT conjecture \cite{Maldacena:1997re}, the holographic description is similar to a 4D strongly interacting theory weakly coupled to some external sources. If the strong dynamics produces a mass gap at a scale of order TeV, there will be a tower of bound states with the lowest excitations at a scale $M\sim$ TeV. This scenario can be mimicked by the 5D model described in the previous paragraph. The truncation corresponding to the case where we retain just the lightest set of bound states.

A 5D realization of the model proposed in this work is given by a gauge theory in a slice of AdS$_5$~\cite{Randall:1999ee}, with the size of the extra dimension of order TeV$^{-1}$. The SU(3)$_c\times$SU(2)$_L\times$SU(2)$_R\times$ U(1)$_X$ bulk gauge theory is broken to the SM SU(3)$_c\times$SU(2)$_L\times$U(1)$_Y$ in the UV by boundary conditions. The 5D fermions have boundary conditions such that their zero-modes are in correspondence with the SM fermions (Neumann boundary conditions on both boundaries (++) for the components with the SM quantum numbers) and they preserve the bulk gauge symmetry in the IR. There is also a 5D field propagating in the bulk, its lightest Kaluza-Klein mode localized towards the IR boundary corresponding to the composite Higgs.~\footnote{One of the simplest realizations consists in a 5D scalar field, but it is also possible to associate the Higgs with the fifth component of a 5D gauge field~\cite{Contino:2003ve,Agashe:2004rs} or with a fermionic condensate~\cite{Burdman:2007sx,Bai:2008gm}.} As usual, the elementary/composite fermionic mixings are controlled by the 5D fermionic mass. The model is fixed by choosing the fermionic embedding under the SU(2)$_R$ symmetry. The different models we have presented in the previous sections correspond to the analogous embeddings of the five dimensional fermions into the SU(2)$_R$ gauge symmetry in the bulk. The mechanism to decouple the top and bottom Yukawa interactions is obtained by introducing two five dimensional fermions associated to $q_L$, their Left-handed components with (++) boundary conditions. The extra zero mode is decoupled by introducing a Right-handed doublet localized in the UV-boundary that marries with a linear combination of the Left-handed fermions through a mass of the order of the UV-scale. As shown in Refs. \cite{Agashe:2004ci,Carena:2006bn,Contino:2006qr}, the fermionic fields with $(-+)$ boundary conditions can give rise to light custodians.

\section{Discussions and Conclusions}\label{conclusions}
The combined measurements of $R_b$ and $A^b_{FB}$ require a large correction in $Zb_R\bar b_R$ and a small correction in $Zb_L\bar b_L$. We have shown that natural models with a new strongly coupled sector that induce EWSB and couple to the SM usually produce large corrections to $Zb_L\bar b_L$. We have also shown how to build a natural model that can simultaneously solve the $A^b_{FB}$ anomaly and reproduce $R_b$, $m_b$ and $m_t$. To keep naturalness in the strongly coupled sector it is necessary to invoke a symmetry ensuring the cancellation of the large contributions to $Zb_L\bar b_L$ arising from the top sector. This condition determines the transformation properties of the top sector under the $O(3)$ custodial symmetry, resulting in two possible embeddings for these fermions \cite{Agashe:2006at} (the custodial symmetry is crucial to protect $\Delta\rho$ from large corrections in composite models of EWSB). To be able to generate the small bottom mass and the corrections to $Zb\bar b$, we have introduced two fermionic resonances, $q^{1cp}$ and $q^{2cp}$, mixing with the SM quark doublet $q_L$, one of them involved in the generation of the top mass and the other one for the bottom mass. The transformation properties of the fermionic resonances that generate the top quark mass are fixed by the symmetry protecting $Zb_L\bar b_L$. The transformation properties of the fermionic resonances that generate the bottom quark mass can be chosen to solve the $A^b_{FB}$ anomaly. We have proposed two different embeddings for the bottom sector, the embedding with larger representations giving rise to better results for $Zb\bar b$, since in this case the corrections to the couplings are parametrically larger and then smaller mixings are required, alleviating the tension with $m_b$. There are larger and more complicated representations that can produce the same effect in $Zb\bar b$, we have considered the smallest and simplest representations that we have found. We have also estimated the size of the one-loop corrections to $Zb\bar b_L$, although a dedicated calculation is needed, our estimates suggest that the tree level contributions dominate.

We have also shown how to obtain the observed $A^b_{FB}$ without a symmetry in the top sector to protect $Zb_L\bar b_L$. To achieve the proper corrections there must be some accidental cancellation between the large contributions from the top sector to $Zb_L\bar b_L$. We have found different models where the corrections arising from the fermionic and bosonic mixings cancel each other. However this cancellation is effective only for some special regions of the parameter space, correlating fundamental parameters that were a priori not related and introducing some tuning in the theory. We have also considered another model, where the contributions from the top sector are induced by the bosonic mixings only, minimizing thus the corrections from this sector. In this case the small correction to $Zb_L\bar b_L$ must be achieved after cancellations with the bottom sector contributions. We have found that in this scenario we can also solve the $A^b_{FB}$ anomaly, but again some tuning is needed to obtain $m_b$ and $R_b$.

As usual in models where the SM masses are generated by mixings with TeV resonances, we have found light fermionic resonances associated to the top sector, with masses of order $0.1-2$ TeV. For one of the two models favoured by naturalness and $A^b_{FB}$, the lightest custodians are lighter than $\sim1$ TeV. These resonances are light compared with the mass scale of the strongly coupled sector, particularly the bosonic mass, that have been set to $2.7-3.4$ TeV. For the models with a symmetry for $Zb_L\bar b_L$, the lightest resonances have charges $Q=5/3$ and $Q=2/3$. We have also found light exotic quarks in the bottom sector, with charges $Q=-4/3$ and $Q=-7/3$, that arise as a consequence of the corrections required for $Zb\bar b$. It seems possible to detect these light fermionic states at the LHC, their observation would be a hint on the solution of the $A^b_{FB}$ anomaly \cite{Contino:2008hi}. We have also found sizable corrections in the $Zt_L\bar t_L$ couplings, of order $-(0.01-0.1)$, that could be measured at the ILC.

We want to mention briefly the heavy Higgs scenario. A heavy Higgs with a mass $m_h\sim 500$ GeV can improve the SM fit to $A^b_{FB}$, almost solving the $A^b_{FB}$ anomaly, without important corrections for $R^b$ \cite{Bouchart:2008vp}. In this case the new physics effects in $A^b_{FB}$ have to be smaller than in the case with a light Higgs, motivating the mechanism to protect $Zb_L\bar b_L$ from large corrections arising from the top sector. However a heavy Higgs is not favoured by other EW precision observables, such as the $S$ parameter. 

Finally, it would be very interesting to build a composite theory of flavour to check if it is possible to solve the $A^b_{FB}$ anomaly in a full flavour theory, reproducing the observed spectrum of quarks and mixings, as well as satisfying the constraints from flavour physics.

In Ref. \cite{Djouadi:2006rk} a model in slice of AdS$_5$ was presented, reproducing the observed $A^b_{FB}$ and $R_b$, but obtaining $m_b\sim 40$~GeV and $m_t\sim 75$~GeV. Note that the deviations are of order $100\%$ for $m_t$ and $m_b$ results one order of magnitude larger than its measured value. The authors argued that either a full theory of flavour or a small hierarchy in the five dimensional Yukawa couplings could solve this problem. In the present work we want to keep naturalness, therefore we have interpreted the wrong spectrum as a signal that the embedding chosen for the fermionic resonances into the SU(2)$_R$ symmetry does not reproduce the observed phenomenology and a better embedding is needed. 

Ref. \cite{Bouchart:2008vp} also proposed a model in a slice of AdS$_5$ that can reproduce the measured spectrum of quarks, $A^b_{FB}$ and $R_b$. The authors introduced embeddings for the fermions under SU(2)$_R$ different from the ones considered in \cite{Djouadi:2006rk} and used a similar mechanism to the one considered here, coupling two composite operators with the elementary doublet $q_L$. The embeddings considered for the composite fermionic operators were different from the ones considered in this work. An important difference with our approach is that, except in one case, the small bottom mass was obtained by introducing a small hierarchy in the fundamental theory, either the five dimensional Yukawa of the bottom sector was small $\lambda_b^{5D} k\simeq0.1-0.2$, with $k$ the natural five dimensional scale, or the bottom mass was suppressed by a small angle in the mixing that leads to the SM doublet $q^{el}_L=(q_L^{1el} \ \sin\gamma+q_L^{2el} \ \cos\gamma)$ (see Eq. (\ref{lmassq12})), with $\cos^2\gamma\simeq 0.01$ (in notation of Ref. \cite{Bouchart:2008vp}, $\cos^2\theta\simeq 0.01$ for their Model II). Ref. \cite{Bouchart:2008vp} also analysed a model very similar to our model $[T1+B1]$, the only difference being that the resonances associated to $t_R$ transform as $({\bf 1},{\bf 3})_{2/3}$, breaking the $P_{LR}$ symmetry. Notably, for this embedding the authors find solutions with natural bottom Yukawa couplings $\lambda_b^{5D} k\simeq1.0-1.7$ and fundamental mixings $\cos^2\gamma\simeq 0.05-0.30$. On the other hand, Ref. \cite{Bouchart:2008vp} claimed, that there does not exist a model with custodial symmetry that can simultaneously reproduce the spectrum of quarks of the third generation and solve the $A^b_{FB}$ anomaly. The authors argue that the reason is that $\delta g^{b_L}$ is systematically too small in this case. We have shown that it is possible to obtain the correct spectrum and to solve the $A^b_{FB}$ anomaly in models with a custodial symmetry by properly choosing the bottom embedding under SU(2)$_R$ (the bottom sector breaks the custodial symmetry protecting $Zb_L\bar b_L$). We found that in general $\delta g^{b_L}$ is small, and for this reason it is necessary to choose a representation for $q^{2cp}$ giving a sizable positive $\delta g^{b_L}$. Even in this case we found some tension between naturalness in the composite sector and $\delta g^{b_L}$, namely a small composite Yukawa $y_{cp}^b$ is preferred if we demand sizable mixings between $b_L$ and $q^{2cp}$ to increase $\delta g^{b_L}$. However, we have shown explicitly that it is possible to find solutions with natural Yukawa couplings $y^{b,t}_{cp}\sim 1$. Moreover, our results show that this tension is alleviated in the models with a custodial symmetry for the top sector, whereas the tension is enlarged if the top sector does not respect any of the $O(3)$ subgroups protecting $Zb_L\bar b_L$.

\bigskip

\noindent
{\bf \large Acknowledgments:} 

We thank Ezequiel \'Alvarez, Alejandro Szynkman and Carlos Wagner for many useful discussions. This work has been supported by the Argentinian CONICET.


\end{document}